\documentclass[times,final]{elsarticle}

\usepackage{jcomp}
\usepackage{framed,multirow}

\usepackage{amssymb}
\usepackage{latexsym}
\usepackage{amsmath}
\usepackage{graphicx}
\usepackage[english]{babel}
\usepackage{subfigure}
\usepackage{caption}
\usepackage[ruled,linesnumbered]{algorithm2e}
\usepackage{bm}

\usepackage{enumitem}

\usepackage{url}
\usepackage{xcolor}
\definecolor{newcolor}{rgb}{.8,.349,.1}

\journal{Journal of Computational Physics}

\begin{document}

\newtheorem{theorem}{Theorem}[section]

\verso{Yiran Meng \textit{etal}}

\begin{frontmatter}

\title{Global physics-informed neural networks (GPINNs): from local point-wise constraint to global nodal association\tnoteref{tnote1}}%

\author{Feng Chen\corref{cor1}}
\author{Yiran Meng}
\author{Kegan Li}
\author{Chaoran Yang}
\author{Jiong Yang}
\cortext[cor1]{Corresponding author: 
  email: 1234@email.com }

\address{School of Electircal Engineering, Xi'an Jiaotong University, Xi'an 710049, China}

\received{}
\finalform{}
\accepted{}
\availableonline{}
\communicated{}

\begin{abstract}
Recently, physics-informed neural networks (PINNs) and their variants have gained significant popularity as a scientific computing method for solving partial differential equations (PDEs), whereas accuracy is still its main shortcoming. 
Despite numerous development efforts, there is no literature demonstrating that these methods surpass classic numerical algorithms in solving the forward issue. 
In this paper, by analyzing the disparities between PINNs and traditional numerical methods based on mesh discretization, we investigate the underlying causes for the inadequate precision of PINNs and introduce a novel approach named global physics-informed neural networks (GPINNs). 
Inspired by the crucial concept of global nodal association in conventional numerical algorithms, GPINNs leverages the prior field distribution information from pre-trained PINNs to estimate the association weights between arbitrary nodes in space. 
GPINNs can not only be regarded as a meshless approach but also be demonstrated, both theoretically and in practical circumstances, to have the ability of second-order convergence when trained with equidistant nodes. 
Overall, GPINNs may be seen as an ideal approach to inheriting the merits of scientific machine learning (SciML) and conventional numerical computing, which also represent the first SciML algorithm to surpass standard numerical methods in terms of accuracy.

\end{abstract}


\end{frontmatter}



\section{Introduction}
Partial differential equations (PDEs) form the foundation of mathematical modeling in the science and engineering.
However, deriving analytical solutions for many problems is challenging except in one-dimensional scenarios and boundary-value problems characterized by regular shapes \cite{Krishnapriyan2021}. 
This difficulty arises because most familiar functions are one-dimensional. 
The advent of computers has facilitated the widespread adoption of numerical techniques, such as regional discretization and piecewise approximation, as the primary methods for solving PDEs in recent decades. 
These methods aim to transform boundary-value problems into a set of positive definite algebraic equations. 
Despite the considerable success of numerical simulations in computational science, these methods require complete information about boundary conditions, medium properties, and excitation as strict criteria. 
Consequently, the resulting nodal data-sets lack generalizability, which significantly limits the applicability of numerical methods in predictive science, such as design, evaluation, and control. 

In recent years, machine learning, especially deep learning, has experienced significant advancements in image recognition and natural language processing \cite{lecun2015,krizhevsky2012,devlin2018}. 
The ability of deep neural networks (DNNs) to approximate high-dimensional function facilitates the 
the seamless integration of large volumes of data. 
Furthermore, DNNs can integrate physical knowledge by designing loss functions, constraint conditions, and optimization algorithm to guide its convergence toward a solution that is consistent with the fundamental physical laws. 
The objective of the data and model fusion-driven strategy is to transform computational science into a comprehensive predictive science, 
thereby establishing a new research domain called scientific machine learning (SciML). 
Physics-informed neural networks (PINNs) and their variants have become the primary representatives of SciML.
Their effectiveness has been demonstrated across various fields, including fluid mechanics, metamaterial design, nano-optics, and electromagnetic wave investigations \citep{raissi2020,chen2020,lu2021}. 
PINNs utilise DNNs as global function approximators for the solution functions of PDEs. 
A loss function is used to ensure that the field distribution at collocation points in the solution domain satisfies the governing equations, 
while the field values at boundary collocation points conform to the given boundary conditions\citep{raissi2019}. 
Since PINNs use the Monte Carlo algorithm to randomly generate uniform collocation points, 
their computational error exhibits grid-independent convergence characteristic. 
This suggests that the computational accuracy of PINNs does not dependent solely on adjusting the number of collocation points. 

The computational error of PINNs can be categorized into two main types: approximation error and generalization error.
The approximation error refers to the discrepancy between the function space represented by the neural network and the solution of the PDEs. 
This error is inherent in the modeling capacities of the neural network and can be reduced by increasing network width and depth. 
 The generalization error, on the other hand, refers to the prediction error of the trained neural network model on a test set. 
This is mainly related to the random sample points used in the discretization of the loss functional. 
A key challenge is that the generalization error may dominate the final error, thus obscuring the modeling capabilities of neural networks. 
To address this, the collocation points in the training set can be adaptive sampled or adaptive weighted according to an appropriate error metrics. 
Specifically, Residual-Based Adaptive Refinement (RAR) techniques and deep adaptive sampling techniques are a class of specific examples. 
These techniques increase sampling density in regions with large PDE residuals or apply importance sampling using probability density function (PDF) 
derived from PDE residuals.\citep{lu2021deepxde,wu2023comprehensive}. 
Another class of approaches to improving the accuracy of PINNs solutions includes methods such as gradient-enhanced physics-informed neural networks (gPINNs) and 
variational physics-informed neural networks (VPINNs). 
These methods can be regarded as adaptive sample-weighted techniques. 
In conclusion, the methods described above reduce the generalization error in approximating the solution to the PDEs by optimizing the neural network model. 
However, when comparing computational errors between classical numerical methods and scientific machine learning in the case of  high-dimensional or parameterized PDEs, 
classical numerical methods often truggle to achieve small approximation errors due to the curse of dimensionality. 
In contrast, PINNs can reduce approximation errors more effective by leveraging deep neural network functions and optimizing the sampling scheme. 
However, for low-dimensional PDEs, 
classical methods such as finite element methods (FEM) circumvent generalization error by using the weak form of the PDEs and the Gaussian quadrature rules. 
Consequently, PINNs are generally less computationally efficient than classical numerical methods due to the prenece of generalization errors.

It remains unclear whether the computational inefficiency of scientific machine learning compared to classic numerical methods is an inherent limitation or untapped potential. 
To further investigate this, a comparative analysis of two representative methods for solving PDEs, PINNs and FEM, can be conducted. 
FEM is based on spatial discrete gride and uses piecewise continuous basis functions and local criteria to compute numerical solutions. 
Thiss is a convex optimization process, 
and  the existence and uniqueness of the solution are provable. 
In contrast, PINNs rely on random sample nodes and is trained by minimizing a pointwise loss functional that satisfies the PDE and adheres to global criteria. 
The parameter optimization process involves balancing the approximation error with the generalization error of the network model. 
This is an inherent where a PDE defined in continuous space necessarily includes the physical information that is missing in the PINNs. 
In general, PDEs contain two principal types of physical information: the field quantifies at nodes and the interactions between nodes in the neighborhood. 
This is analogous to Kirchhoff's law for node voltage, node current, and the volt-ampere relationship between neighboring nodes. 
In conclusion, the FEM scheme incorporates both types of physical information at the nodes and between neighboring nodes, allowing it to find a unique solution under the definite conditions. 
In contrast, the PINNs' loss function employs a pointwise approximation strategy that neglects the correlation between neighboring collocation points. 
Not withstanding the continuous updating of collocation point positions, this strategy fails to eliminate the generalization error introduced by overfitting between the collocation points.

To this end, we propose a novel Global Physics-Informed Neural Networks (GPINNs) in the paper. 
By continuously learning the interactions between nodes during training, the GPINNs are able to provide a high-fidelity solutions to PDEs based on random sampling. 
Specifically, 
the GPINNs' loss function is formulated as a system of algebraic equations, consisting primarily of topological operators and constitutive relations. 
In the global solution domain, the topological operators for the node network is defined containing only the elements 1, -1, and 0, 
which reflect the adjacency relationships between nodes and are independent of their specific positions. 
In the local solution domain, the DNNs predicted field distribution is used to accurately recover the local constitutive relations and determine the interactions between adjacent nodes. 
The concept of solving differential equations with nodal networks can be traced back to the equivalent magnetic network method in the 1960s and has gradually developed into the discrete geometric approaches based on finite integral technique and the cell method based on geometric topology. 
Therefore, we aim to combine classical numerical methods with DNNs-based PDE solvers, and propose a node network-based SciML approach to achieve high-accuracy solutions at low computational cost. 

The structure of this paper is organized as follows: After the introduction in section 1, 
section 2 presents a comprehensive examination of the generalization error of PINNs.
Then we offer a thorough overview of the method for nodal topology network construction and the expression of constitutive relationship between physical quantities. 
Next, we analysis the disparities between PINNs and conventional numerical methods in solving PDEs.
Finally, a novel method called GPINNs is put out, which transitions from local point-wise constraint to global nodal association. 
In addition, in section 3, we provide the mathematical proof for the convergence of the proposed technique and arrange a 1D numerical experiment to demonstrate it. 
Section 4 provides the solution results under various nodal samplings to further demonstrate the efficacy of the GPINNs. 
In Section 5, a PDE with the domain shaped like a map of Shaanxi Province is solved to show how well the new approach can handle challenging boundary problems. 
Section 6 discusses the more complex problem of fragmented uniform media, for which the proposed approach similarly provides a solution with a higher accuracy. 
Lastly, we provide a summary of our findings in section 7.

\section{Random Node Network-Based GPINNs}

It is not comprehensive to analyze PINNs solely from the perspective of computational errors.
Additional study is required for understanding the distinctions between the physical properties of PINNs and the original physical problem.
Taking the continuity equation commonly found in physics as an example, 
it is a differential equation that describes the transport behavior of conserved quantities such as mass, energy, momentum, and charge.
Any continuity equation can be represented either as a differential operator form applicable at a specific point or as a flux integral form applicable to any finite region.
Among these, the integral form is the local version of the conservation law, 
which means that the change in a certain conserved quantity within any region is equal to the amount flowing into or out of the region from its boundary.
The model error of PINNs is precisely because it disrupts the local conservation of the original problem.

\begin{figure}[!t]
  \centering
  \includegraphics[scale=.9]{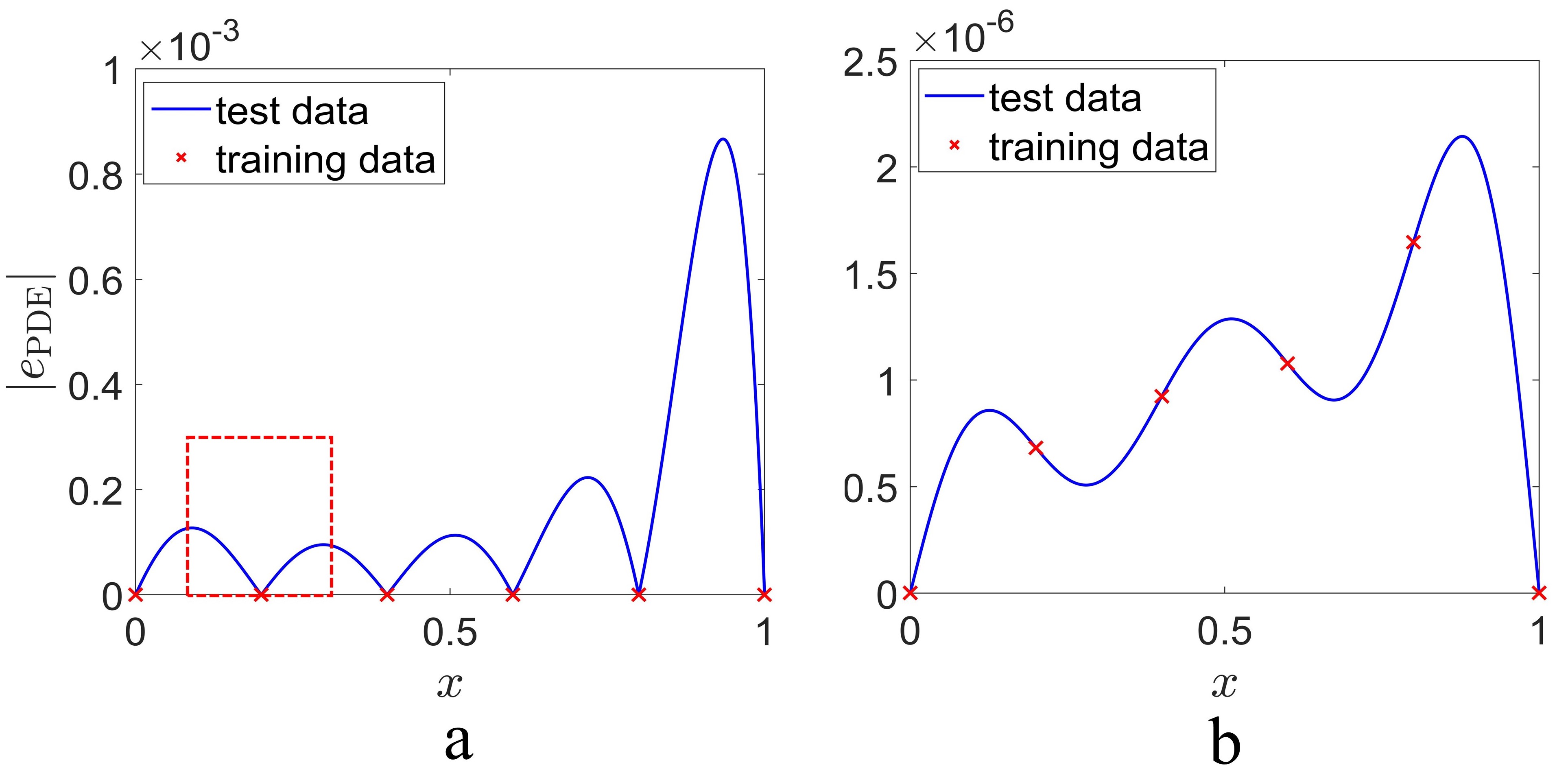}
  \caption{The computational error of PINNs:
  (a) Absolute residual of PINNs.
  (b) Absolute computational error in approximating the PDE solution using PINNs.}
\end{figure}

Taking the one-dimensional continuity equation as an example, its PDE can be expressed in the following form:
\begin{equation}
  \begin{aligned}
    -&\nabla\cdot \left(e^x\nabla u\left(x\right)\right)=f(x)\quad x\in\begin{pmatrix}0,1
    \end{pmatrix}\\
    &u\left(0\right)=0\\
    &u\left(1\right)=\cos\left(1\right)
  \end{aligned}
\end{equation}

where $f(x)=-e^x\left[\cos(x)-2\sin(x)-x\cos(x)-x\sin(x)\right]$. The analytical solution to the problem is $ x\cos(x)$.
PINNs uses a neural network function $u_\theta $ with the structure of [1,20,1] to approximate the solution function of the PDE. Tanh is chosen as the activation function, where $\theta$ represents the parameters of the network to be trained.
PINNs uses the loss function to ensure that the continuity Equ. 1 is satisfied at four equidistant configuration points within the domain $(0,\,1)$, 
while the field quantities at the boundary configuration points $u_\theta (x_i) (i=1,6)$ meet the given boundary conditions.
The absolute value of the PDE residuals of PINNs is shown in Fig. 1(a). 
In the red region of the figure $[0.1,0.3]$, the PDE residual is approximately zero only at the point $x=0.2$, 
while other areas exhibit overfitting with relatively large PDE residuals. 
When the integral form of the continuity equation $\oint _{\delta \mathbf{V}}-e^x\nabla  u_\theta (x)\cdot d\overline{S} = \int _\mathbf{V} f(x)\,d\mathbf{V}$ is applied to the finite space $[0.1,0.3 ]$, 
a source or sink caused by the computation exists on the boundary surface, 
leading to the emergence of spurious solutions and an increase in overall computational error.
The computational error of the PINNs approximating the PDE solution is shown in Fig. 1(b). 
Whether at training points or test points, the non-conservative PINNs method cannot provide relatively accurate computational results.

\subsection{Node Network Representation Learning}

Unlike the non-conservative PINNs method based on random point sampling, 
this paper proposes the GPINNs method based on a conservative node network.
The core idea of GPINNs is to change the loss function from local pointwise constraints to global node-related network constraints.
The node network mainly studies the different elements represented by the nodes and the connections between these elements. 
This model can not only ensure the uniqueness of the solution but also, due to the its conservativeness, can provide high-precision solutions.

Using randomly configured points to generate node networks not only allows for a natural expression of the connection relationships between nodes 
but also retains the unique advantages of the Monte Carlo method in solving certain specific problems.
Particularly, for the first reason, the computational complexity of random node networks mainly depends on the number of sample nodes, which is independent of the dimension, 
whereas the computational complexity of conventional numerical methods based on mesh discretization increases exponentially with the dimensions.
Therefore, this method is very effective in solving certain high-dimensional problems. 
Second, random node networks are useful for tackling some optimization issues, 
particularly global optimization problems, 
since they offer a flexible and effective global search approach through random sampling and the large sample simulation.
This is also the reason why neural network training usually employs random sampling.

There are various methods for generating node networks using randomly configured points, including the Delaunay triangulation generation algorithm, 
the minimum spanning tree algorithm, the K-nearest neighbors algorithm, and so on. 
In order to reduce the number of edges to connecting points, thereby saving storage and computational costs, 
this paper adopts the Delaunay triangulation, an efficient geometric sparse structure. 
Fig. 2(a) displays the Delaunay triangulation that was generated using 10 randomly arranged points. 
This node network has locality, meaning that adding, deleting, or moving a node will only affect the triangular network adjacent to it. 
Moreover, by considering the circumsphere, the Delaunay triangulation generation method can also be extended to three-dimensional or higher dimensions; 
in addition to Euclidean distance, other metrics can also be generalized. 

Dual networks are typically created in order to extract complementary information from the prime network, 
such as analyzing the connecting relationships between nodes in the prime network. 
The Barycentric dual network algorithm and the Voronoi dual network algorithm are the two primary techniques for creating dual networks. 
To improve the simplicity and generality of the algorithm, this paper adopts the Barycentric dual network as a complementary topological structure, as illustrated in Fig. 2(b).

\begin{figure}[!t]
  \centering
  \includegraphics[scale=1]{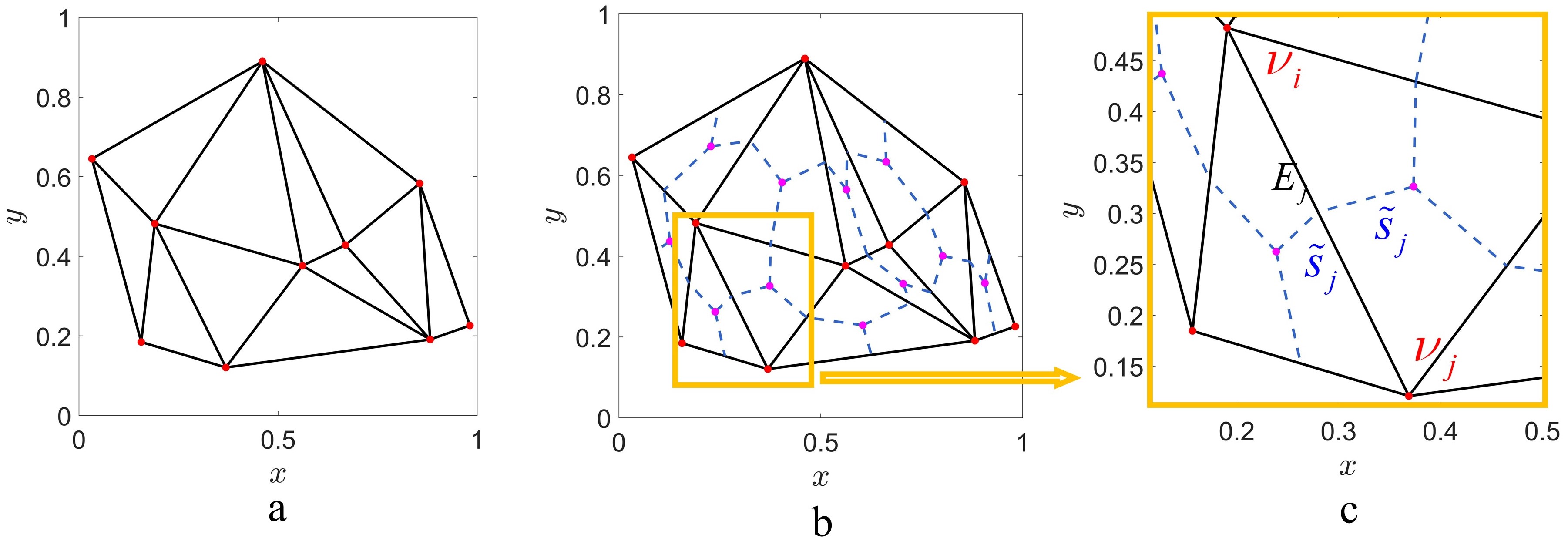}
  \caption{Construction of Random Node Networks:
  (a) Generation of Delaunay Triangulation Network.
  (b) Generation of Barycentric Dual Network.
  (c) Local Magnification of Node Network.}
\end{figure}

Network representation is the bridge that connects the raw data with network application tasks. 
How to describe the network data in a reasonable way is the initial challenge for the node-network based GPINNs algorithm.
Analogous to the general description ways of topological equations and constitutive relationships in traditional numerical methods like the cell method, 
partial differential equations for different physical fields can be transformed into algebraic equations based on primal-dual networks. 
Taking the Laplace operator, which is common in partial differential equations, as an example, 
the differential operator defined in continuous space can be transformed into an algebraic equation on a random node network, which can be expressed as

\begin{equation}
  \begin{aligned}
    \nabla\cdot \left(\varepsilon \nabla u\right)\Rightarrow \mathbf{G^\top MGU} 
  \end{aligned}
\end{equation}
It involves some basic concepts of network representation, defined as follows:
\begin{itemize}[label=\textbullet]
  \item $\bm{\mathit{V}}$ (vertex node): A node is a functional entity in a network. An important application of network representation learning is the prediction of node labels $\bm{U}$.
  \item $\bm{\mathit{E}}$ (edge):Edges are used to depict the relationship between two adjacent nodes. In addition to the label information attached to the nodes, there is also a wealth of interaction information between the nodes on the edges, such as constitutive relationships.
  \item $\bm{\mathit{N}}$ (network): A network is used to depict relational data. Define the network $\bm{N}=(\bm{V},\bm{E}) $, where $\bm{V}$ is the set of nodes of $\bm{N}$, and $\bm{E}$ is the set of edges of $\bm{N}$.
  \item $\bm{G}$ (The Incidence Matrix):The incidence matrix $\bm{G}$ corresponding to the network $\bm{N}\in(\bm{V},\bm{E} ) $ is of size  $| \bm{E}\vert \times |\bm{V}\vert  $. 
  The incidence matrix directly provides the matrix representation of the network topology. As shown in Fig. 2, if $(v_i,v_j )\in \bm{E} $, then $\bm{G}_{ij}=1$ or -1; otherwise, $\bm{G}_{ij}=0$.
  \item $\bm{M}$ (edge feature matrix): The edge feature matrix $\bm{M}$ corresponding to the network $\bm{N}\in(\bm{V},\bm{E} ) $ is of size $| \bm{E}\vert \times |\bm{E}\vert  $. 
  $\bm{M}$ is a highly sparse matrix, with non-zero elements only on the diagonal, used to represent the direct link relationships between network nodes, that is: 
  \begin{equation}
    \begin{aligned}
      M_{jj}=\frac{\Phi _j}{\Delta U _j}=\frac{\int_{\mathit{s} _j}\varepsilon \nabla U_\theta \cdot \mathrm{d}\widetilde{\bm{\mathit{s}} }}{\int_{E_j}\nabla U_\theta \cdot \mathrm{d}\bm{\mathit{E}} }
    \end{aligned}
  \end{equation}
  where $\Phi_j$ and $\Delta U_j$ are the extensive quantity flow and intensive quantity difference defined on the prime network edge $\mathbf{E_j}$ and the dual network face $\widetilde{\bm{\mathit{s}}_j}$, respectively.
  The spatial relationship between the original edge $\mathbf{E_j}$ and the dual face $\widetilde{\bm{\mathit{s}}_j}$ is shown in Fig. 2(c). 
  $U_\theta$ is the neural network used to approximate the PDE solution function, where $\theta$ represents the network parameters to be trained.
\end{itemize}

Next, let's take the Poisson equation as an example to explain in detail how to use the GPINNs method to solve partial differential equations, whose general form is: 
\begin{equation}
  \begin{aligned}
    \nabla\cdot \left(\varepsilon \nabla u\right) - f(\bm{x} )=0  \quad \bm{x}\in\Omega 
  \end{aligned}
\end{equation}
Where \( U \) represents the solution of the PDE, and \( \Omega \) is a subset of \( \mathbb{R}^n \). We define the node network equation corresponding to the PDE in Equ. (4) as:
\begin{equation}
  \begin{aligned}
    \bm{f}:=\vec{\mathbf{G}}^\top\vec{\mathbf{M}}(\theta,\vec{E},\vec{S})\vec{\mathbf{G}}\vec{\mathbf{U}}(\vec{x}) -\bm{\mathit{F} } 
  \end{aligned}
\end{equation}
Where $\vec{\mathbf{F} } $ is the excitation source vector, and its elements are $\mathbf{F} _i=\int f(\vec{x} )d\tilde{\mathbf{V} _i} $, $\tilde{\mathbf{V}} _i$ is the dual cell corresponding to the node $\tilde{x} _i$. 
The assumption of approximating U using neural networks, together with Equ. (5), constitutes the GPINNs (Global Physics-Informed Neural Networks). 
This network has the same parameters $\theta$ as the network representing U and can learn the shared parameters between the neural networks $U_\theta$ and $\tilde{f}$ by minimizing the mean squared error loss. 

\begin{equation}
  \begin{aligned}
    Loss( \bm{x},\theta)=\frac{1}{N_f}\sum_{i = 1}^{N_f} |\tilde{\bm{f}}({\bm{x}}_i  )\vert ^2 + \frac{1}{N_\upsilon }\sum_{j = 1}^{N_\upsilon} |u_\theta ({\bm{x}}_j) - {\bm{U}}_j\vert ^2
  \end{aligned}
\end{equation}
Where \( u_j\vert _{j=1}^{N_{\upsilon}} \) represents the boundary training data on \( U(\vec{x} _j) \), and \( x_i\vert _{i=1}^{N_f} \) refers to the configuration points of \( \tilde{f}(\vec{x}) \).

Although scientific machine learning methods like PINNs are convenient and straightforward, 
they are often affected by the problem of computational efficiency. 
On one hand, PINNs disrupt the physical principal of the original problem, such as local conservation. 
And the computational errors which PINNs introduce can sometimes be unacceptable compared to traditional numerical methods. 
On the other hand, PINNs alter the existence and uniqueness conditions of the solution of the original boundary value problem, 
and this non-uniqueness of the solutions makes neural network training difficult. 
In contrast, GPINNs learn representations of the original boundary value issue using random node networks. 
Node network representation learning can be either semi-supervised or unsupervised. 
The node network representation of the physical field and the neural network representation of the solution function can be automatically obtained by optimization algorithms. 
The conservative node network representation enables precise and effective algorithm construction. 

\subsection{Algorithm Workflow of GPINNs} 
GPINNs use prior information of physical field distribution to learn node network representations.
Methods for obtaining prior field distributions include scientific machine learning methods such as PINNs, Deep Ritz, XPINNs, Deep Galerkin, and Deeponet.
At this stage, it is not necessary to calculate the prior field distribution exactly, but merely to roughly approximate it.
As a result, compared to traditional scientific machine learning techniques, a significantly fewer number of training iterations are needed to acquire prior knowledge about the field distribution. 
For example, in the case of Equ. (1), the PINNs training iteration steps are 40,000, the deep Ritz method requires 50,000 steps, while the prior field distribution prediction estimation only needs about 5,000 steps.
\begin{algorithm}[t]
  \caption{Algorithm of GPINNs}
  Specify the network architecture, initialize network parameters, and set training hyperparameters.\;
  Apply the PINNs algorithm to obtain prior field distribution information. \;
  Generate a topological network with $N_n$ random nodes.\;
  Construct the corresponding node adjacency matrix $\mathbf{G}$. \;
  Use the preliminarily trained neural network model to predict the edge feature matrix $\mathbf{M}$. \;
  Formulate the flux conservation loss function (see Equ. (6)).\;
  Perform gradient-based optimization to find the optimal neural network parameters that minimize the loss function from Step 6

\end{algorithm}

In order to acquire previous knowledge of the field distribution, this article employs random sampling and pointwise satisfaction of the PDE loss functional.
Introducing this prior knowledge into the GPINNs method has three advantages:

\begin{itemize}[label=\textbullet]
  \item Algorithm consistency. This includes maintaining consistency between random sampling points in the physical field and topological network nodes;
  ensuring the neural network model of pointwise approximation of PDEs is consistent with the conservation-type node network; 
  and maintaining consistency in the algorithm for differentiating the input coordinates of the neural network.
  \item Algorithm feasibility. The prior and posterior knowledge of neural network models can be derived using the chain rule and automatic differentiation.
  \item Initial value effectiveness. The gradient of the physical field distribution serves as the initial value for the GPINNs method. 
  As shown in Fig. 3, the computational errors of solving the boundary value problem (1) using FEM and PINNs are compared. 
  Although the prediction error of the solution function U by PINNs is relatively large, the prediction accuracy of its solution gradient is very high. 
  This is mainly because PINNs uses continuous differentiable activation functions and automatic differentiation techniques.
\end{itemize}

\begin{figure}[!h]
  \centering
  \includegraphics[scale=.9]{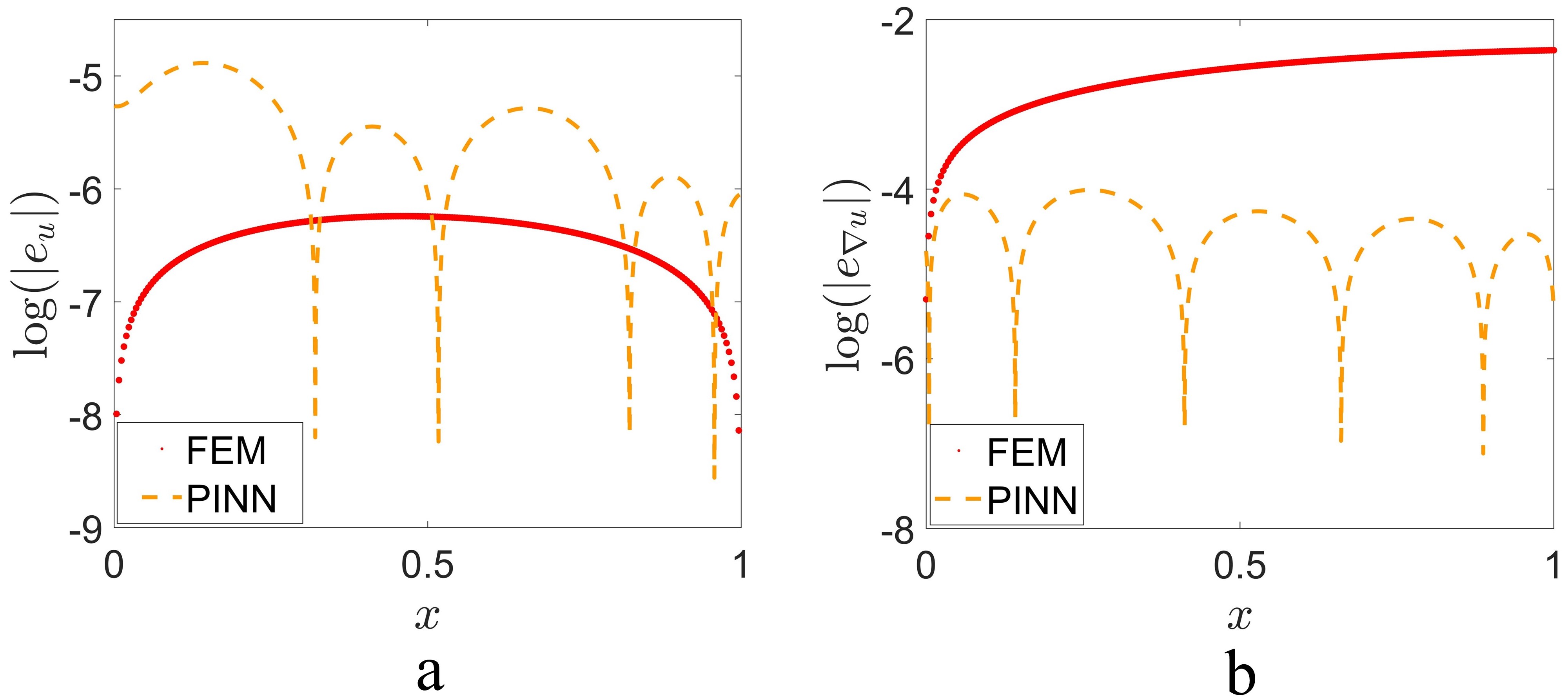}
  \caption{Comparison of Errors Between PINNs and FEM with 256 Equidistant Nodes:
  (a) Error comparison of the solution function.
  (b) Error comparison of the solution function gradient.}
\end{figure}
The specific algorithm implementation of GPINNs can be referenced in the pseudocode of Algorithm 1, which mainly consists of the following six steps:

\begin{itemize}[label=\textbullet]
  \item Step 1: Determine the width, depth, and activation functions of the neural network structure. Initialize weights, biases, and other network parameters, and set the learning rate and iteration steps as hyperparameters for network training.
  \item Step 2: Use random sampling and pointwise satisfaction of the PDE loss functional to train the neural network, obtaining prior information of the field distribution.
  \item Step 3: Use random sampling points and the Delaunay algorithm to generate the node topology network.
  \item Step 4: Generate the incidence matrix $\vec{\mathbf{G}}$ based on the topological relationship of the node network; generate the edge feature matrix $\vec{\mathbf{M}}$ based on Equ. (3) and the prior field distribution.
  \item Step 5: Construct the loss function based on the conservation-type node network according to Equ. (6).
  \item Step 6: Use the ADAM optimizer to adjust the neural network parameters to satisfy the conservation loss function in Step 5.
\end{itemize}

\section{Convergence analysis of GPINNs}

A class of scientific machine learning methods represented by PINNs, 
due to the use of a loss function that satisfies the PDE pointwise, 
leads to the destruction of the physical properties of the original problem, 
thereby introducing model errors, such as local conservation.
At this point, even if PINNs increase the number of iterations and sampling points, it cannot effectively improve the training results.
We first consider the numerical example of Equ. (1), as shown in Fig. 4, 
the predictive error are plotted with respect to number of training data points from 1 to 1000. 
Despite the fact that PINNs' result converge relatively quickly in relation to the quantity of training data, 
we find that the prediction error constantly stays at the $10^{-5}$ level and that the error saturates over 10 data points. 
By analogy to the convergence analysis of PINNs in reference \citep{shin2020convergence}, 
we will derive the convergence mode of GPINNs.

\subsection{Mathematical Model and Assumptions}

Let \( U \) be a bounded domain in \( \mathbb{R}^d \), and \( \Gamma \) be the boundary of the domain. We consider the form of the differential equation as:
\begin{equation}
  \begin{aligned}
    \mathcal{L} [u](\mathbf{x})&=f(\mathbf{x}),\quad\forall\mathbf{x}\in U,\\
    \quad\mathcal{B}[u](\mathbf{x})&=g(\mathbf{x}),\quad\forall\mathbf{x}\in\Gamma\subseteq\partial U
  \end{aligned}
\end{equation}

Among them, $\mathcal{L}[\bullet]$ is a differential operator, and $\mathcal{B}[\bullet]$ can be Dirichlet, Neumann, Robin, or periodic boundary conditions.For simplicity, we only consider the case of Dirichlet boundary conditions. $u$ is the solution to PDEs (1).

Before analyzing the convergence of GPINNs, we make the following assumptions about the training data distribution, the operator $\mathcal{L}[\bullet]$, and the function $f$:
\begin{itemize}[label=\textbullet]
  \item 1. For $\varepsilon >0$, there exists partitions of U, denoted as $\{ \widetilde{A}_{x_r^j}\}_{j=1}^n  $, which depend on
  depend on $\varepsilon $, such that each of these dual cells $ \widetilde{A}_{x_r^j} $ has a side length of $\varepsilon $ and its centroid is located at $x_r^j$. 
  \item 2. There exists positive constant $C_r$, such that $\forall \varepsilon $, the partitions from
  the above satisfy
  \begin{equation}
    \begin{aligned}
      C_r\varepsilon ^d\leq \mu _r(A_{x_r}^j)
    \end{aligned}
  \end{equation}
  where $C_r$ depend only on $( U,\mu _r) $, 
  and $\mu _r$ is probability distributions defined on \(U\). The empirical probability distribution is defined as
  \begin{equation}
    \begin{aligned}
      \mu _r^n = \frac{1}{n}\sum_{j = 1}^{n}\delta _{x_r^j}
    \end{aligned}
  \end{equation}
  The neural network is trained using random sampling and a pointwise PDE residual loss functional to obtain prior information on the field distribution.
  \item 3. Let the operator $\mathcal{L}[\bullet]$ and the function \(f\) be uniformly Hölder continuous in \(U\) with exponent $\alpha $, 
  Taking \(f\) as an example, this can be expressed as:
  \begin{equation}
    \begin{aligned}
      \left[f\right]_{\alpha _j, U}=\sup _{x,y\in U, x\neq y} \frac{|f(x)-f(y) \vert }{\|x-y \Vert ^\alpha  } <\infty , 0<\alpha \leq 1
    \end{aligned}
  \end{equation}
\end{itemize}

\begin{figure}[!t]
  \centering
  \includegraphics[scale=0.06]{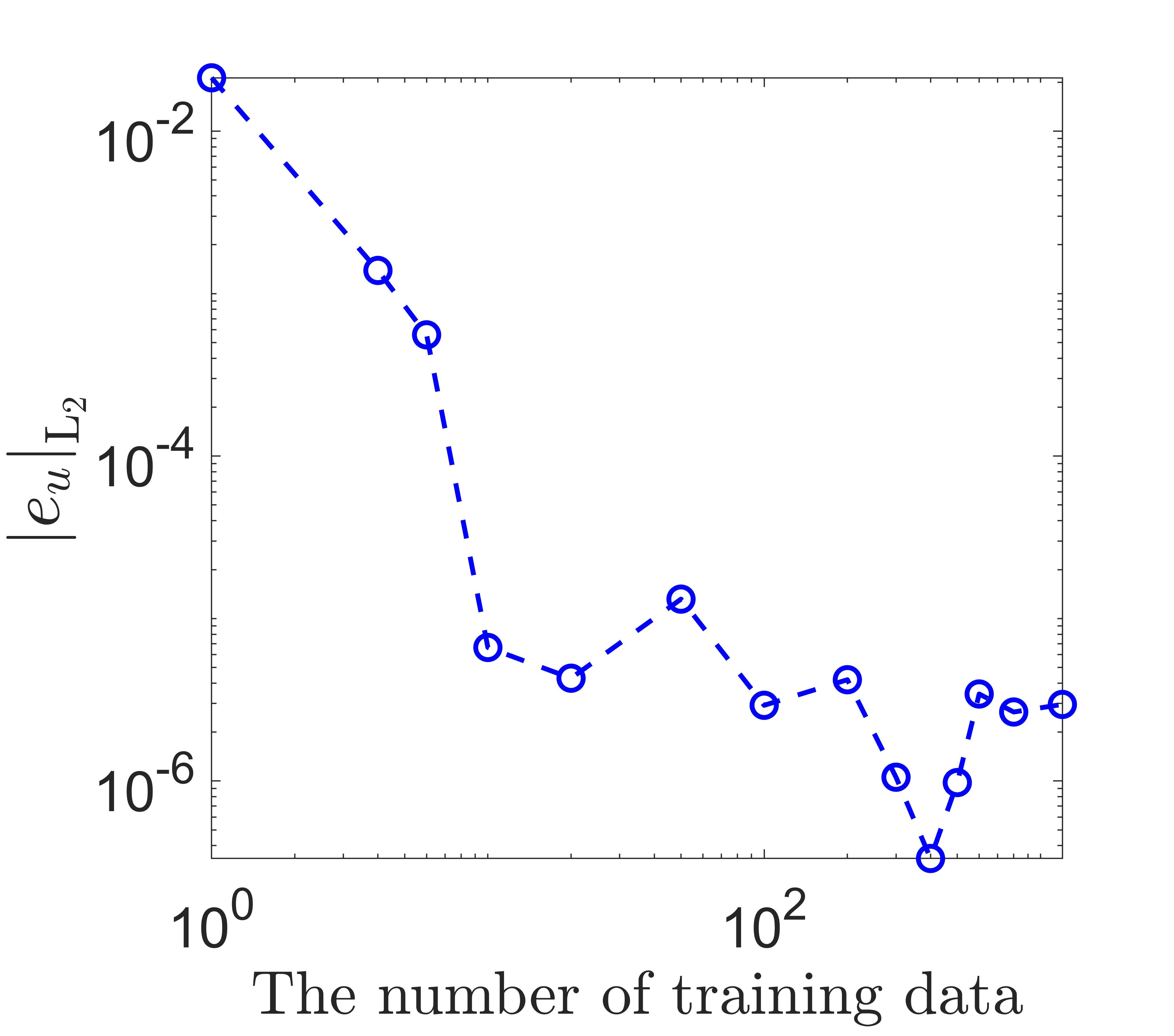}
  \caption{\(L_2\) Error of PINN Under Different Training Points}
\end{figure}

Given a class of neural networks $H_m$ that depend on the number of training samples $n$, we will seek to find a neural network $h^*$ in $H_m$ that minimizes the objective function. 
if the expected loss function were available, a function that minimizes it would be sought. 
However, since the expected loss is not attainable in practice, an empirical loss function is adopted. 
The expected loss of GPINNs, $Loss^{\mathrm{GPINNs}}(h)$, and the empirical loss, $Loss_n^{\mathrm{GPINNs}}(h)$, can be expressed as:
\begin{equation}
  \begin{aligned}
    Loss^{\mathrm{GPINNs}} = \sum_{j = 1}^{n}  \|\int _{A_{x_r^j}}[\mathcal{L} [h](x_r)-f(x_r)]d\nu  \Vert ^2
  \end{aligned}
\end{equation}
\begin{equation}
  \begin{aligned}
    Loss_n^{\mathrm{GPINNs}} = \sum_{j = 1}^{n}  \|\int _{A_{x_r^j}}[\mathcal{L} [h](x_r^j)-f(x_r^j)]d\nu  \Vert ^2
  \end{aligned}
\end{equation}

\subsection{Convergence Analysis}

We first derive the upper bound of the expected GPINNs loss (11), which includes a specific regularization empirical loss. According to the mean value theorem for integrals, Equ. (11) can be expressed as:
\begin{equation}
  \begin{aligned}
    Loss^{\mathrm{GPINNs}} = \sum_{j = 1}^{n}  \|\int _{A_{x_r^j}}[\mathcal{L} [h](x_{\xi}^j)-f(x_{\xi}^j)]d\nu  \Vert ^2
  \end{aligned}
\end{equation}
Where \( x_\xi^j \) is the integral midpoint of \( A_{x_r^j} \).

Based on the inequality $\|x+y+z \Vert^2\leq 3(\| x\Vert^2+\| y\Vert ^2+\| z\Vert ^2 ) $, we obtain
\begin{equation}
  \begin{aligned}
    \sum_{j = 1}^{n}\|\int _{A_{x_r^j}} [\mathcal{L}[h](x_{\xi}^j)-f(x_{\xi}^j)]\Vert ^2  
    \leq &3\sum_{j = 1}^{n}\{ \|\int _{A_{x_i}^j}[\mathcal{L}[h](x_{\xi}^j)-\mathcal{L}[h](x_r^j)]d\nu  \Vert^2 \\
    &+\|\int _{A_{x_r}^j}[f(x_r^j)-f(x_{\xi}^j)]d\nu \Vert ^2 +
    |\int_{A_{x_r}^j}[\mathcal{L}[h](x_r^j)-f(x_r^j)]d\nu \Vert ^2\}
  \end{aligned}
\end{equation}

The distance between any two points in the dual cell is less than $\sqrt{d}\varepsilon$. That is, $\|x_{\xi}^j - x_r^j\Vert \leq \sqrt{d}\varepsilon$. According to the definition of Hölder continuity, inequality (14) is expressed as:
\begin{equation}
  \begin{aligned}
    Loss^{\mathrm{GPINNs}}(h) \leq  3d^{\alpha }\varepsilon_r ^{2\alpha}[\mathcal{L}[h]]^2_{\alpha;U}\varepsilon_r^{2d}+3d^{\alpha }\varepsilon_r ^{2\alpha}[f]^2_{\alpha;U}\varepsilon_r^{2d}+3Loss_n^{GPINN}(h)
  \end{aligned}
\end{equation}
By letting
\begin{equation}
  \begin{aligned}
    n=\frac{1}{C_r^2\varepsilon_r ^{2d}}
  \end{aligned}
\end{equation}
we have 
\begin{equation}
  \begin{aligned}
    Loss^{\mathrm{GPINNs}}(h) \leq  3[\mathcal{L}[h]]^2_{\alpha;U}d^{\alpha }C_r^{-\frac{2\alpha+2d}{d}}n^{-\frac{\alpha+d}{d}}
    +3[f]^2_{\alpha;U}d^{\alpha }C_r^{-\frac{2\alpha+2d}{d}}n^{-\frac{\alpha+d}{d}}
    +3Loss_n^{\mathrm{GPINNs}}(h)
  \end{aligned}
\end{equation}

Assuming that the neural network possesses sufficient expressive capability and is well-trained, we can neglect both the approximation error and optimization error. Consequently, the empirical loss can be approximated as
\begin{equation}
  \begin{aligned}
     \text{Loss}_n^{\mathrm{GPINNs}}(h)=0 
  \end{aligned}
\end{equation}
Under this assumption, we can derive the convergence behavior of GPINNs as follows:
\begin{equation}
  \begin{aligned}
    Loss^{\mathrm{GPINNs}}(h) =\mathcal{O} (n^{-\frac{\alpha}{d}-1})
  \end{aligned}
\end{equation}

In the context of machine learning, this is equivalent to the convergence of generalization error. 
Compared to the convergence characteristics of the expected loss of PINNs $Loss^{\mathrm{PINNs}}(h)=\mathcal{O} (n^{-\frac{\alpha}{d}})$, 
the generalization error of GPINNs converges at a rate that is an order of magnitude faster.

\section{Numerical validation}

\begin{figure}[!t]
  \centering
  \includegraphics[scale=0.8]{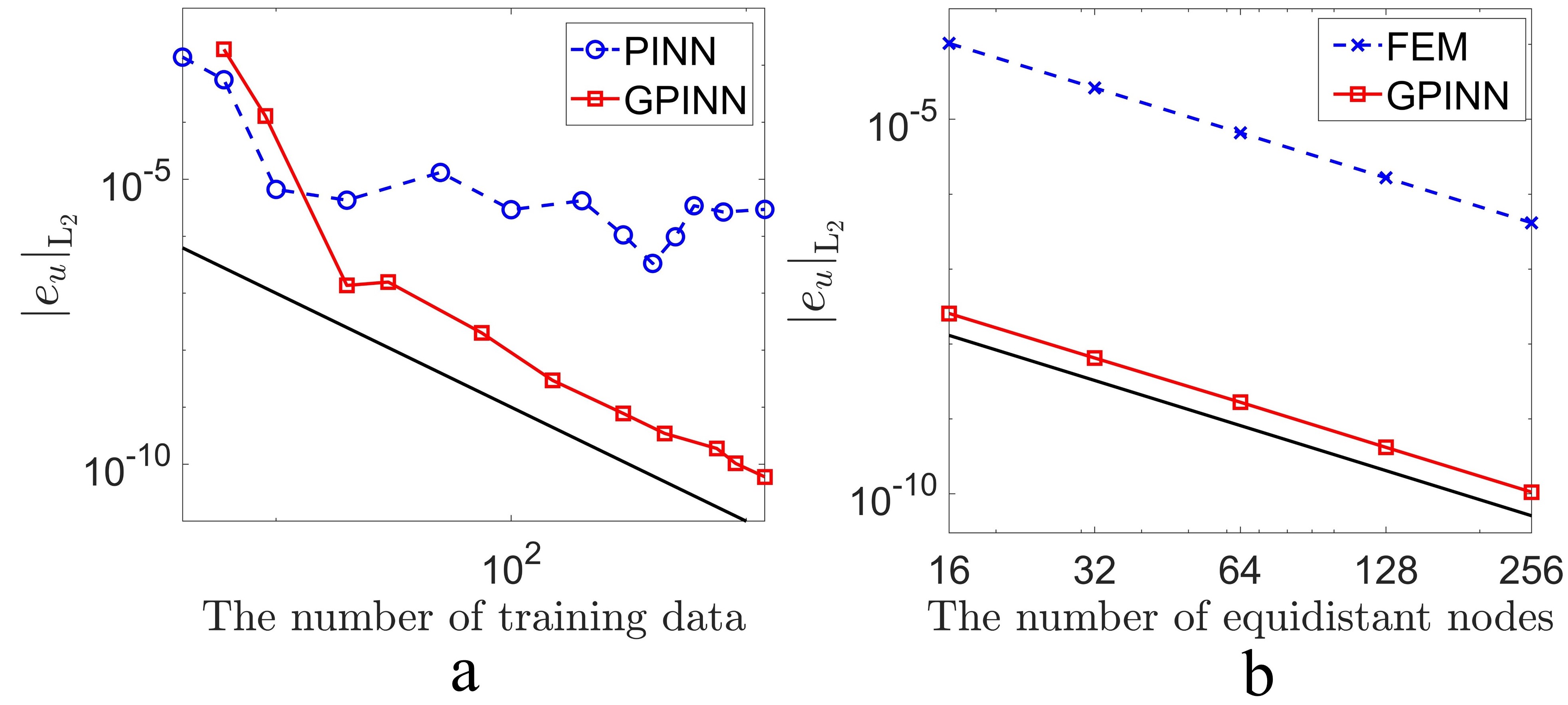}
  \caption{ Convergence Comparison of PINNs, GPINNs, and FEM for the 1D Poisson Problem:
  (a) Convergence results of PINNs and GPINNs at different sampling points.
  (b) Convergence comparison between GPINNs with different topological nodes and FEM.}
\end{figure}

This section uses GPINNs to solve multiple partial differential equation cases to verify the superiority of this method in terms of generalization error convergence and computational accuracy. 
First, we take a one-dimensional model as an example and analyze the convergence of FEM, PINNs, and GPINNs through uniform step-size partitioning or sampling. 
Second, we have confirmed the robustness and flexibility of GPINNs in boundary handling for PDE situations with curved and complicated boundaries. 
Finally, we provide the results of GPINNs, highlighting its advantages in handling discontinuous media problems, which is difficult for standard PINNs model. 
We compared the results of four different test cases, which provided the maximum absolute error and the $L_2$ norm between the numerical solution (predicted solution) $U_n'$ and the exact solution $U_n$, namely: 
\begin{equation}
  \begin{aligned}
    |e_u \vert_{\max}=\max _n|U_n-U_n' \vert 
  \end{aligned}
\end{equation}
\begin{equation}
  \begin{aligned}
    |e_u \vert_{L_2}=\sqrt{\frac{\sum_{n = 1}^{N} (U_n-U_n')^2 }{N}}
  \end{aligned}
\end{equation}
Where N is the number of sampling points. We mentioned that all numerical experiments in this paper are implemented based on the Pytorch framework and completed on a computing platform configured with an NVIDIA 3090ti graphics card.
\subsection{Comparison and Analysis of Convergence}
In order to compare and examine the convergence characteristics of PINNs, GPINNs, and FEM, we consider the one-dimensional Poisson equation in Equ. (1) of Section 2 of this paper, where the computational domain \([0,\,1]\) is sampled or partitioned with equal step sizes. 
PINNs and GPINNs use a fully connected network with a structure of \([1,\,50,\,50,\,1]\), which consists of an output layer with one neuron, two hidden layers with 50 neurons each, and an output layer with one neuron. 
Using the Adam algorithm for loss function optimization, the total number of iterations is 50,000 steps, with a learning rate of 1e-4. 
Unlike the training process of PINNs, GPINNs reallocates sampling points based on different utilization ways. 
The 3/4 of the sampling points are used for pointwise constraints of the governing equations to gain previous knowledge of the field distribution, while the remaining 1/4 of the sampling points are used to create a node topology network for flux conservation constraints. 
The calculation results are shown in Fig. 5(a). For reference, a convergence line of $\mathcal{O} (n^{-2})$ is added to the figure, represented by a solid black line. 
Consistent with the error analysis in Section 3, GPINNs converges at a rate of $\mathcal{O} (n^{-2})$. 
The convergence curve of PINNs shows a saturation trend after the number of sampling points reaches 10. 
Continuing to increase the number of sampling points not only increases the complexity of the loss function, affecting computational efficiency, but also brings negligible improvements in accuracy. 
Therefore, we use the field distribution information obtained from these 10 sampling points as the initial values for GPINNs to generate network. 
The 16, 32, 64, and 128 equidistant nodes in the interval \([0,\,1]\) are used as GPINNs topological nodes and solved. 
At the same time, the results obtained by the FEM at the same nodes are used for comparison, where FEM uses piecewise linear functions as shape functions.
Fig. 5(b) illustrates that while FEM may also attain second-order convergence speed, GPINNs outperforms FEM in terms of overall accuracy. 
GPINNs uses merely 10 sampling points to obtain prior knowledge, which results in a $10^3$ to $10^4$ improve in accuracy.

\subsection{Hollow Circle Problem}
We now focus on the effectiveness of GPINNs in handling curved boundaries and compare it with PINNs and FEM. 
Consider a two-dimensional Laplace equation, where the computational domain is a square with side length 2, with a circular region of radius 0.5 removed from the center, as shown in Fig. 6. 
The analytical solution can be easily obtained using the method of separation of variables: 
\begin{figure}[!t]
  \centering
  \includegraphics[scale=.78]{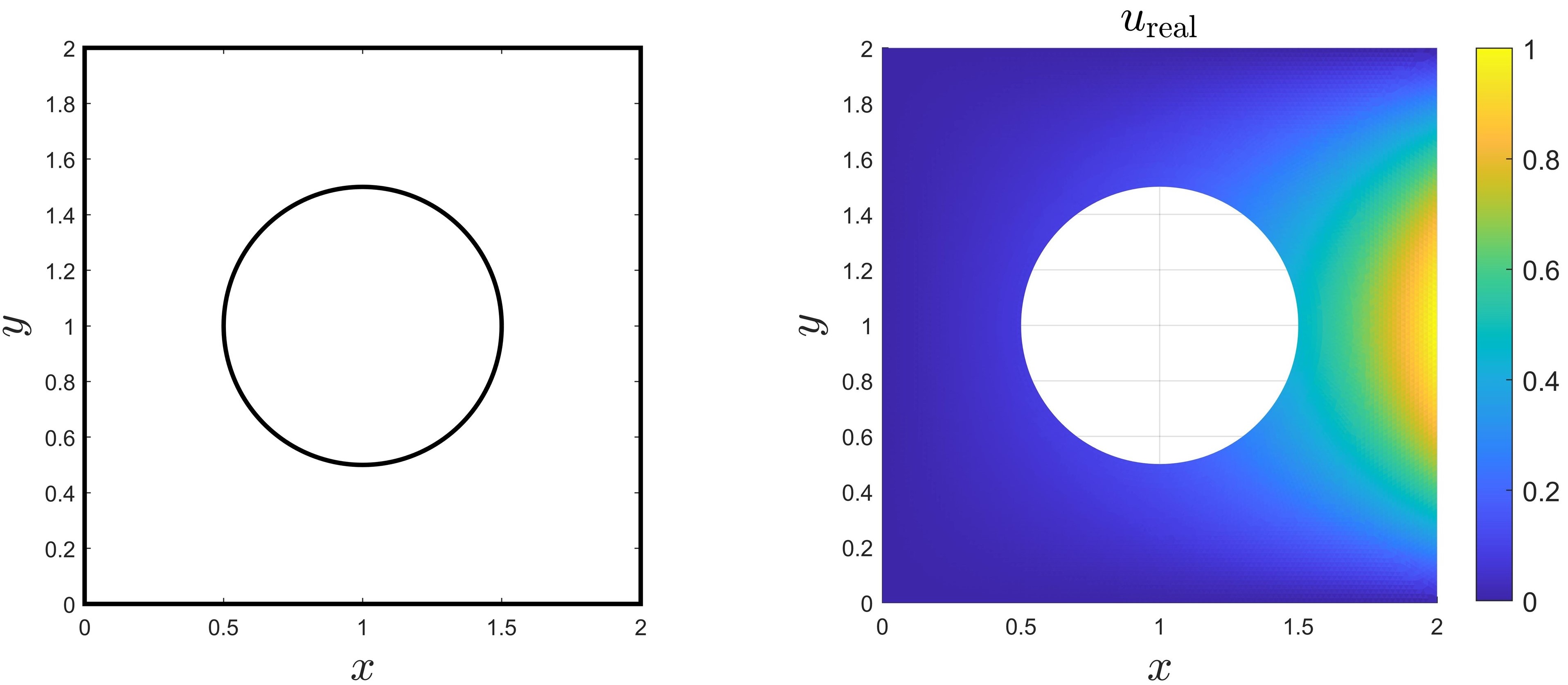}
  \caption{Example in Section 4.1: the left part shows the computation domain and the right part shows the analytical solution.}
\end{figure}

\begin{equation}
  \begin{aligned}
    U(x,y)=\frac{sinh(\pi x/2)sin(\pi y/2)}{sinh(\pi)}
  \end{aligned}
\end{equation}
The boundary conditions here are $U\vert _{x=0}=U\vert _{y=0}=U\vert _{y=2}=0$, and $U\vert _{x=2}=sin(\pi y/2)$.

\begin{figure}[!h]
  \centering
  \includegraphics[scale=.78]{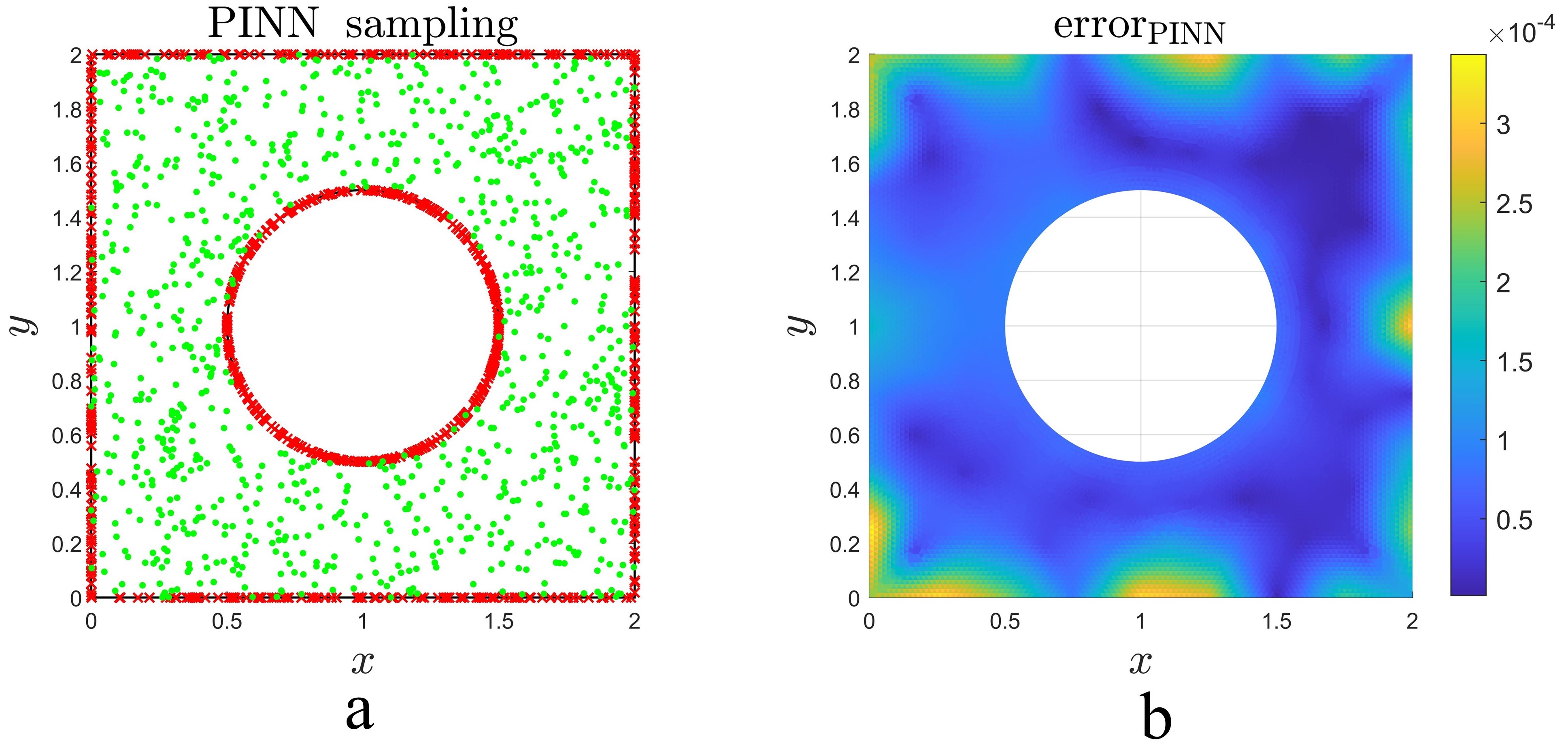}
  \caption{ Example in Section 4.1 solved by PINNs: (a) shows scatter points randomly sampled at each epoch from the boundary marked by the red crosses and interior marked by the green points. (b) shows the absolute value error of PINNs.}
\end{figure}

First, we use PINNs to solve this partial differential equation, where the structure of the neural network is designed as \([2,\,50,\,50,\,50,\,50,\,1]\), which includes an input layer with two neurons, four hidden layers with 50 neurons each, and an output layer with one neuron. 
During the training iterations, we randomly sampled 800 and 1000 configuration points from the boundaries and the interior of the domain, respectively, for node constraints. 
The spatial distribution of the training points is shown in Fig. 7(a), where the boundary configuration is marked by red crosses, and the internal configuration points are represented by green circles. 
The total number of iterations for network training is set to 30,000 steps, while other hyperparameters (such as learning rate, decay rate, and activation function) remain consistent with the previous settings. 
The prediction results of PINNs are shown in Fig. 7(b).The maximum absolute error of PINNs is $3.48\times 10^{-4}$, while the $L_2$ norm error is $1.27\times 10^{-4}$.
However, there are notable error oscillations in the predicted solution within the local range close to the boundary. 
This phenomenon can be attributed to the inconsistent convergence of the gradient flow of the residuals of the PDEs and the boundary condition residuals during the training process. 
For an in-depth theoretical analysis and numerical experimental validation of such issues, further reference can be made to the literature \citep{wang2021understanding}.

Next, we use FEM to solve the PDE problem. 
Since the model includes curved boundaries, traditional triangulated element-based discretization methods struggle to accurately represent its geometry, thereby introducing significant geometric errors during the numerical solution process. 
This limitation mainly stems from the insufficient ability of triangular elements to approximate curved boundaries, resulting in a deviation between the discretized geometric representation and the actual boundary. 
To reduce such errors, higher density meshes or curved edge elements are usually employed. 
It will, however, greatly increase computational complexity and could result in numerical problems like mesh production difficulties or decreased algebraic system solving efficiency. 
Consequently, one of the major challenges that classical numerical approaches encounter is solving curved boundary problems.

\begin{figure}[!h]
  \centering
  \includegraphics[scale=.68]{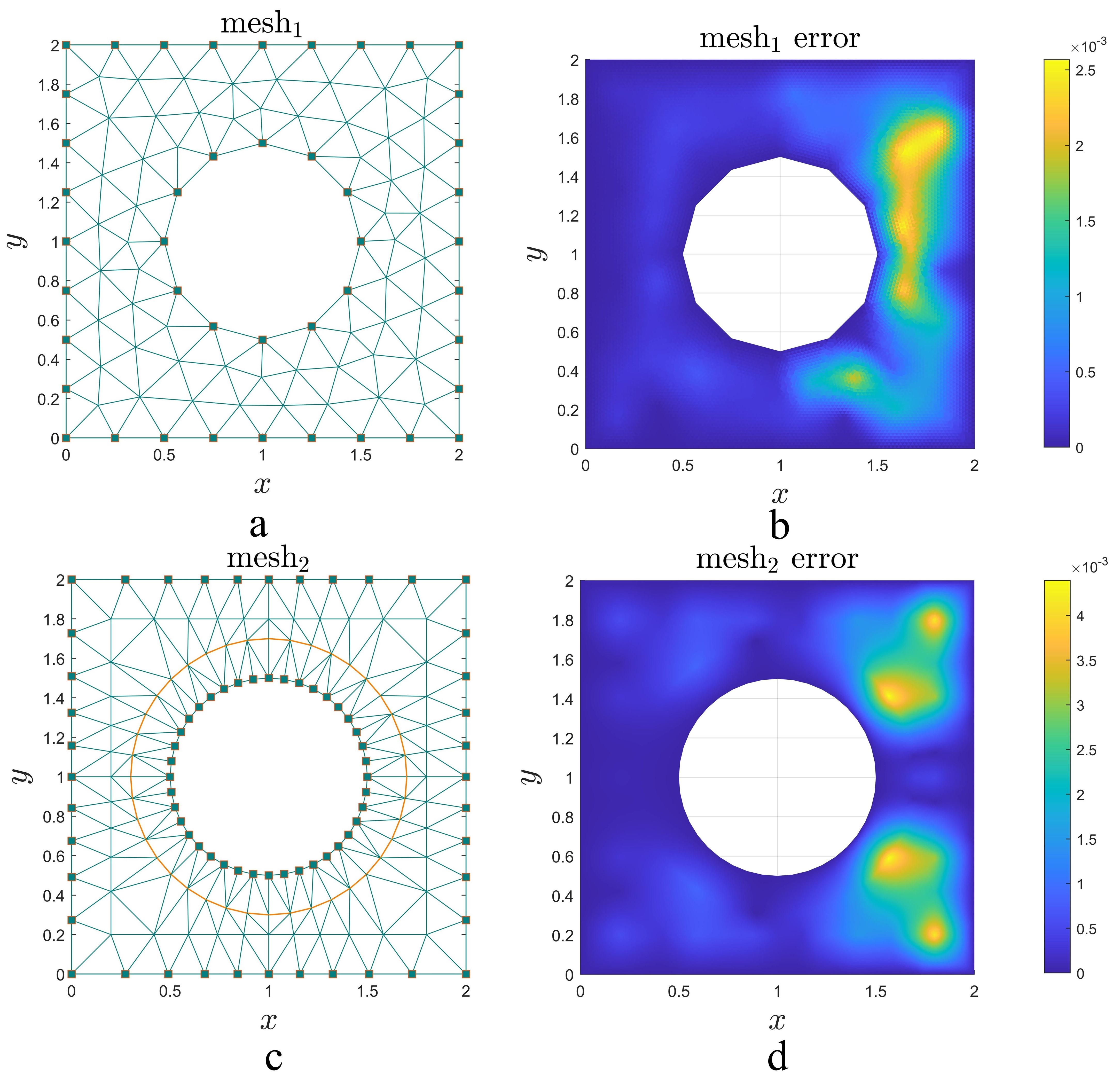}
  \caption{Estimation Results of FEM Under Different Discrete Elements:
  (a) Discretization of Mesh 1.
  (b) Computational error under Mesh 1 discretization.
  (c) Discretization of Mesh 2, where the orange line represents an additionally generated curved boundary.
  (d) Computational error under Mesh 2 discretization.}
\end{figure}

  In Mesh 1, we use Delaunay triangulation to discretize the solution domain, which contains 44 boundary points and 66 interior points, resulting in 156 discrete elements. 
  In Mesh 2, there are 80 boundary points and 80 interior points, resulting in a total of 240 discrete elements. 
  Fig. 8 shows two different types of mesh discretizations and their corresponding FEM results. 
  Due to the fact that the computational accuracy of traditional numerical algorithms highly depends on mesh quality, which is often referred to as mesh regularity. 
  This can be evaluated through a basic metric, which is the minimum ratio of the circumradius to the inradius of all discrete triangular elements. The calculation formula is as follows: 

\begin{equation}
  \text{Regularity}=\inf\left\{\frac{r_1}{R_1},\frac{r_2}{R_2},\frac{r_3}{R_3},...,\frac{r_n}{R_n}\right\}
\end{equation}

  where $n$ represents the total number of units, and $r$ and $R$ represent the inradius and circumradius of the unit, respectively. 
  According to the Equ. 23, the regularity of Mesh 2 is 0.278, while the regularity of Mesh 1 is 0.460. 
  The maximum absolute error and $L_2$ norm error of FEM on the two grids are summarized in Table 1. 
  Despite having more elements than Mesh 1, Mesh 2's maximum absolute error and $L_2$ norm error have both increased noticeably. 
  This is a typical manifestation of the grid dependency of traditional numerical methods.

\begin{table*}[!h]
  \caption{\label{tab1}Results of node-based traditional numerical method under different types of meshes}
  \centering
  \begin{tabular}{p{1.5cm}p{3.0cm}p{2.2cm}p{4.6cm}p{2.85cm}}
    \hline
   Type & Number of elements & Regularity & Maximum absolute value error & $L_2$ norm error\\
  \hline
  Mesh 1 & 156 & 0.460 & 2.61e-3 &7.40e-4\\
  Mesh 2 & 240 & 0.278 & 4.42e-3 &1.04e-3\\
  \hline
  \end{tabular}
\end{table*}

Finally, we use GPINNs to solve this partial differential equation problem. 
The sampling configuration points of GPINNs are maintained constant with the discrete node positions in FEM in order to guarantee the validity of the comparative experiment. 
The node topology networks generated using coarse sampling and local refined sampling are shown in Fig. 9(a) and 9(c), referred to as Network 1 and Network 2, respectively. 
The node networks generated using the Delaunay algorithm and their dual networks generated using the barycentric algorithm are represented by dashed lines and solid lines, respectively. 

Unlike traditional numerical methods that approximate curves with piecewise linear segments and solve PDEs in simplified domains, 
GPINNs can accurately describe the geometric features of the solution domain by introducing prior field distributions, thereby avoiding geometric errors. 
Here, we directly use the training results of PINNs as prior knowledge to estimate the edge features of the node network. 
However, PINNs' error fluctuations in the boundary regions affect the uniformity of the overall error distribution and introduce noise into the gradient estimation of the field distribution. 
As a result, the accuracy improvement in the 2D situation is an order of magnitude lower than the $10^3\sim 10^4$ times improvement in the 1D scenario.
Fig. 9(b)(d) show the pointwise absolute error results of GPINNs on two networks, indicating an accuracy improvement of 2 to 3 orders of magnitude. 
Table 2 presents detailed information from the result analysis. 
Notably, Network 2's $L_2$ error is lower than Network 1's, which contrasts significantly with the results of the conventional numerical techniques. 
This is because GPINNs is a meshless method, which is a significant advantage compared to traditional numerical methods. 
Discrete meshes are typically used in traditional numerical methods to approximate the solution domain, 
and the accuracy and stability of computations are directly impacted by the mesh's quality. 
In contrast, GPINNs avoid the errors introduced by geometric generation and geometric approximation by using neural networks to directly learn the solution function of PDEs. 
Furthermore, complex geometries can be handled by GPINNs naturally without the need for extra mesh preparation. 
This greatly improves the method's applicability and robustness in complex problems while also simplifying the preprocessing step.

\begin{figure}[!t]
  \centering
  \includegraphics[scale=.68]{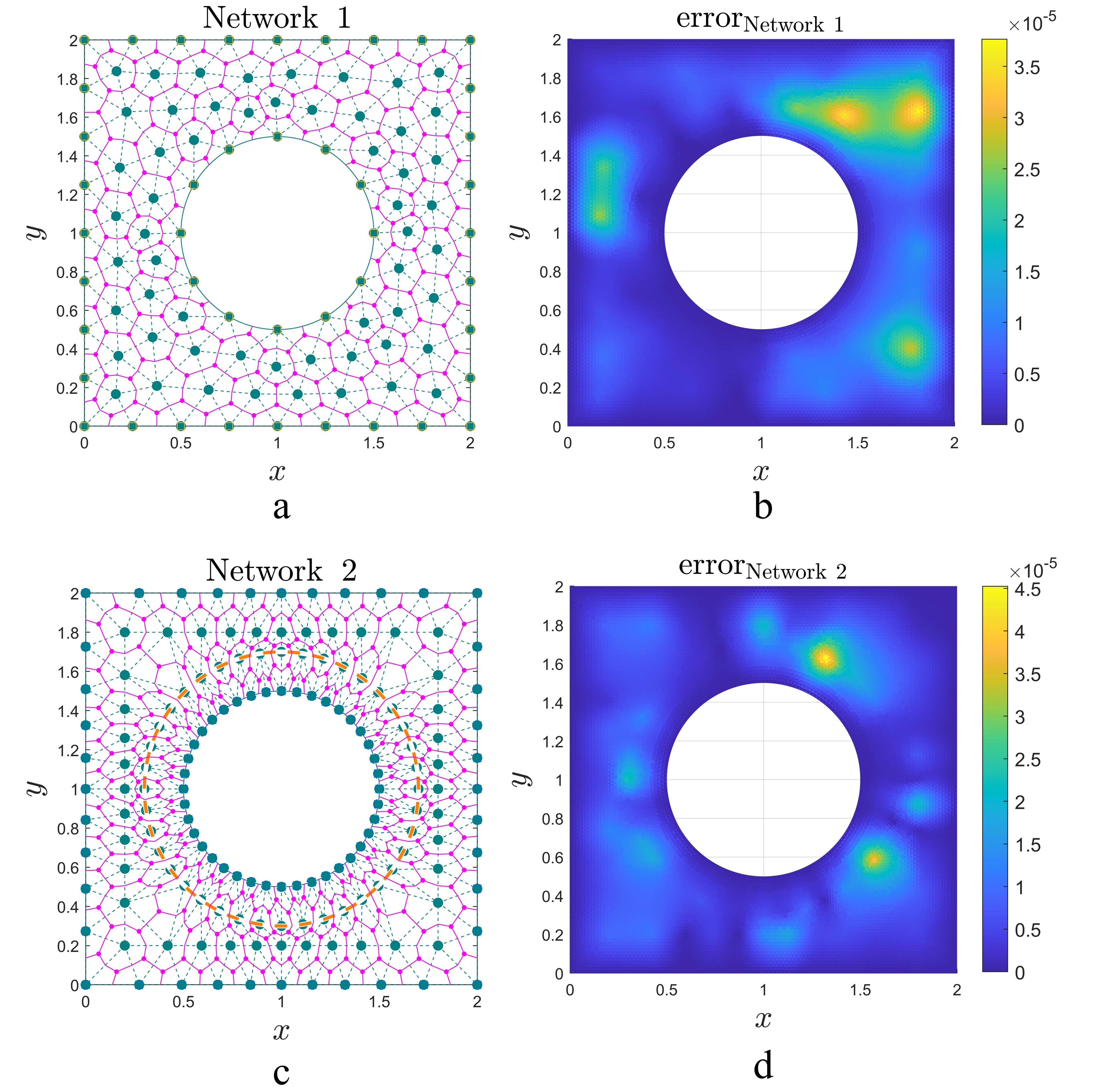}
  \caption{Computation Results of GPINNs Under Different Topological Networks:
  (a) Network 1 corresponding to Mesh 1.
  (b) Absolute error of GPINNs under Network 1.
  (c) Network 2 corresponding to Mesh 2.
  (d) Absolute error of GPINNs under Network 2.
}
\end{figure}

 \begin{table*}[!h]
  \caption{\label{tab1} Results of GPINNs under Association 1, 2 and 3}
  \centering
  \begin{tabular}{p{3.0cm}p{5.cm}p{2.4cm}p{4.cm}p{4.cm}}
    \hline
   Type  & Maximum absolute value error & $\mathrm{L}_2$ norm error\\
  \hline
  Network 1 & 3.80e-05 &9.58e-06\\
  Network 2 & 4.81e-05 &8.65e-06\\
  \hline
  \end{tabular}
\end{table*}
\begin{figure}[!t]
  \centering
  \includegraphics[scale=.7]{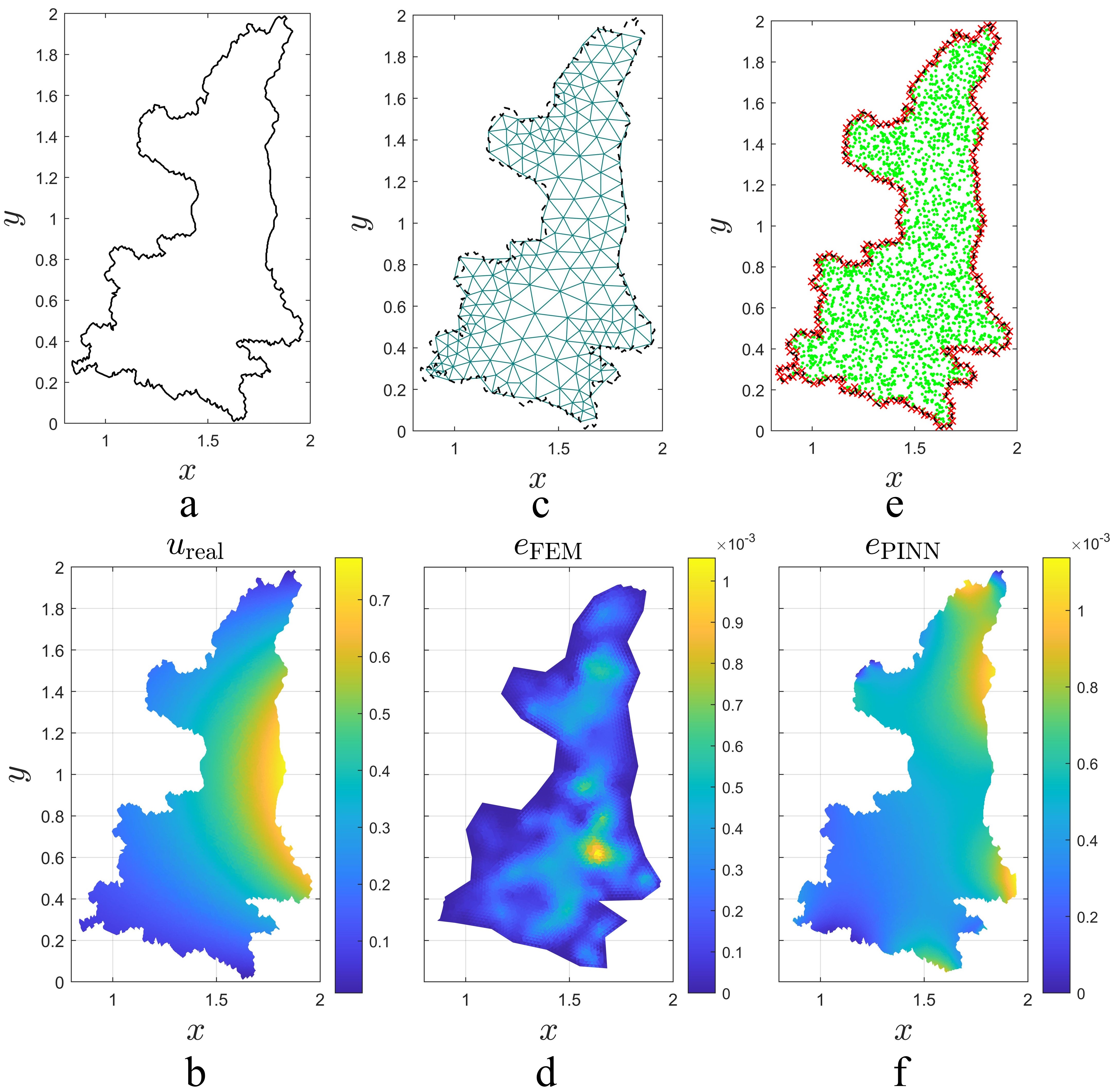}
  \caption{Real solution of example in Section 4.2 together with absolute value error of node-basd numerical method and PINNs. 
  (a) The map of Xi'an city is used as computation domain. 
  (b) shows the real solution.
  (c) Domain discretization using Delaunay triangles. The domain boundaries are simplified, which the black dashed lines refer to the real boundaries. 
  (d) The absolute value error of FEM. 
  (e) shows scatter points randomly sampled at each epoch from the boundary marked by the red crosses and interior marked by the green points. 
  (f) shows the absolute value error of PINNs.}
\end{figure}
\subsection{Complex Boundary Problem}
The case in this section still uses the Laplace equation as the governing equation and introduces highly complex boundary conditions to verify the robustness of the method. 
The solution area is based on the map of Xi'an as the outline, with its boundary composed of nearly 1200 polylines, as shown in Fig. 10(a). 
The analytical solution of the equation remains unchanged, expressed as $sinh(\pi x/2)sin(\pi y/2)/sinh(\pi)$, and this analytical solution is shown in Fig. 10(b). 
Dirichlet boundary conditions can be derived from the analytical solution. 
The boundary geometric features of such problems are extremely complex, which poses significant challenges to traditional numerical methods. 
Traditional numerical techniques discretize the solution domain using triangular elements, as seen in Fig. 10(c). 
However, triangular components are inherently limited in their ability to describe complicated geometric boundaries. 
In order to accurately depict the geometric features of the boundary, 
traditional methods need to rely on mesh refinement or a large number of low-regularity triangular elements. 
This not only causes the mesh generation process to be time-consuming or even fail, but also affects the solution accuracy. 
Therefore, traditional methods can only solve the problem by simplifying the boundaries. 
However, this introduces geometric errors, and implementing automation of the simplification process is challenging, necessitating a large amount of manual intervention. 
As shown in Fig. 10(c), the actual geometric boundary is indicated by the dashed line, while the simplified area is represented by the solid green line. 
This simplification obviously leaves out a great deal of border information. 
Although the FEM results can be guaranteed, as shown in Fig. 10(d), 
the results are only limited to the domain of the simplified solution.

Compared to traditional numerical methods based on discrete triangular elements, random sampling offers a high degree of freedom in describing complex geometric domains, which is one of the main advantages of the PINNs algorithm. 
Fig. 10(e) shows the distribution of random sampling points used during the PINNs training process, with boundary points marked by red crosses and interior points represented by green circles. 
In this case, the total number of boundary points is 2600, and the number of interior points is 2200. 
The neural network structure is adjusted to \([2,\,80,\,80,\,80,\,80,\,1]\), which means an input layer with two neurons, four hidden layers with 80 neurons each, and an output layer with one neuron. 
The other hyperparameters remain consistent with before. 
The pointwise absolute error distribution of the PINNs is displayed in Fig. 10(f). The error features, which include greater fluctuations close to the boundaries, are in line with previously noted occurrences.

\begin{figure}[!t]
  \centering
  \includegraphics[scale=.65]{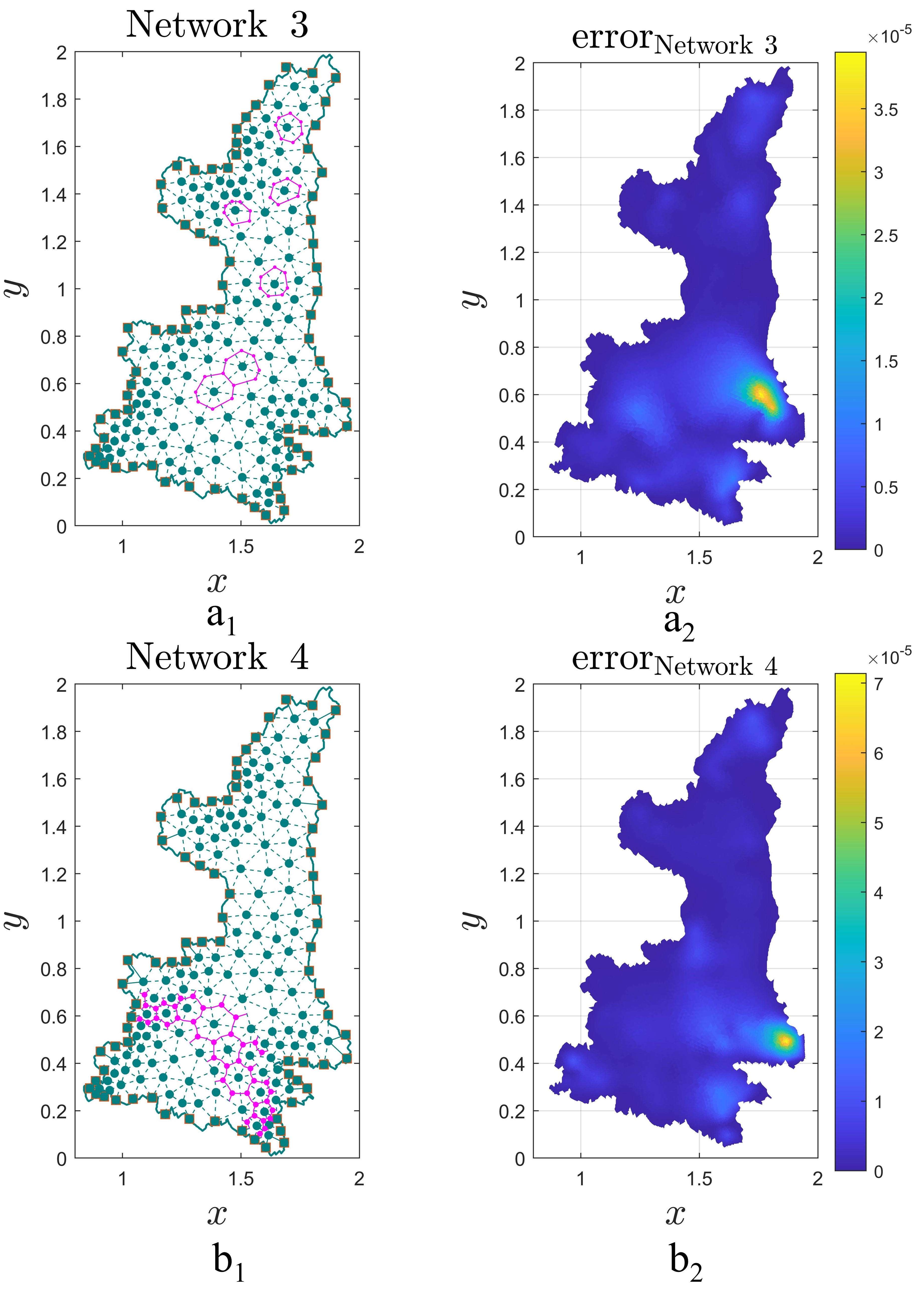}
  \caption{Computation Results of GPINNs in Network 3 and Network 4:
  (a1) Network 3 corresponding to the Voronoi dual cell complex.
  (a2) Absolute error of GPINNs under Network 3.
  (b1) Network 4 corresponding to the barycentric dual cell complex.
  (b2) Absolute error of GPINNs under Network 4.}
\end{figure}

Finally, we use GPINNs to solve this problem. 
GPINNs construct topological networks through random nodes, without the need for any geometric simplification of the solution domain. 
As in the previous case, we use the same topological nodes but adopt two different forms of dual network generation. 
Network 3 uses the circumcenter of the triangle (the Voronoi algorithm) as the dual network vertex, while Network 4 uses the barycentric of the triangle (the barycentric algorithm) as the dual network vertex. 
From Fig. 11, it can be seen that GPINN achieved higher accuracy on both Network 3 and Network 4. 
Because of the orthogonality of its dual network, Network 3 showed improved accuracy recovery. 
Table 3 details the prediction errors of FEM, PINN, and GPINN at the same nodes.

\begin{table*}[!h]
  \caption{\label{tab1} Results of the different solving methods}
  \centering
  \begin{tabular}{p{3cm}p{4cm}p{5.2cm}p{3cm}}
    \hline
   Type & Number of nodes & Maximum absolute error & $L_2$ norm error \\
  \hline
  FEM     & 198 & 1.11e-03   &2.19e-04 \\
  PINNs   & 198 & 1.14e-03   &5.19e-04 \\
  GPINNs Network 3 & 198 & 4.09e-05   &4.63e-06 \\
  GPINNs Network 4 & 198 & 7.28e-05   &7.28e-06 \\
  \hline
  \end{tabular}
\end{table*}
\section{Piecewise Homogeneous Media Problem}

Although a large number of studies have focused on improving the accuracy and solving speed of PINNs \citep{jagtap2020,kharazmi2021,yu2022gradient,pang2019fpinns}, existing methods still face significant challenges in dealing with problems involving piecewise homogeneous media. 
In this section, we will tackle this problem by building a conservation-type node topology network, 
drawing on the piecewise neural network approach of XPINNs and the idea of flux conservation of cPINNs.
We adopt the Xi'an Jiaotong University School of Electrical Engineering's logo as the solution domain, as seen in Fig. 12-a. 
Three distinct sub-regions, each with unique medium properties, make up this emblem.
According to Fig. 12-b, 
a material having a dielectric constant of 1 fills Domain 1 (shown in green) and a material with a dielectric constant of 3 fills Domain 2 (shown in blue). 
A substance having a dielectric constant of 0.2 fills Domain 3 (shown in red).
The entire emblem shape's computational domain has a uniform distribution of charge density. 
The potential at the circular region's outer boundary is zero since it satisfies the homogeneous boundary criterion. 
The electrostatic field's governing equation is as follows:

\begin{equation}\begin{cases}
  -\nabla\cdot(1\cdot\nabla V)=6 & (x,y)\in\mathrm{Domain}1 \\
  -\nabla\cdot(3\cdot\nabla V)=6 & (x,y)\in\mathrm{Domain}2 \\
  -\nabla\cdot(0.2\cdot\nabla V)=6 & (x,y)\in\mathrm{Domain}3 \\
  V=0 & (x,y)\in\partial\Omega 
  \end{cases}\end{equation}

  \begin{figure}[!t]
    \centering
    \includegraphics[scale=.7]{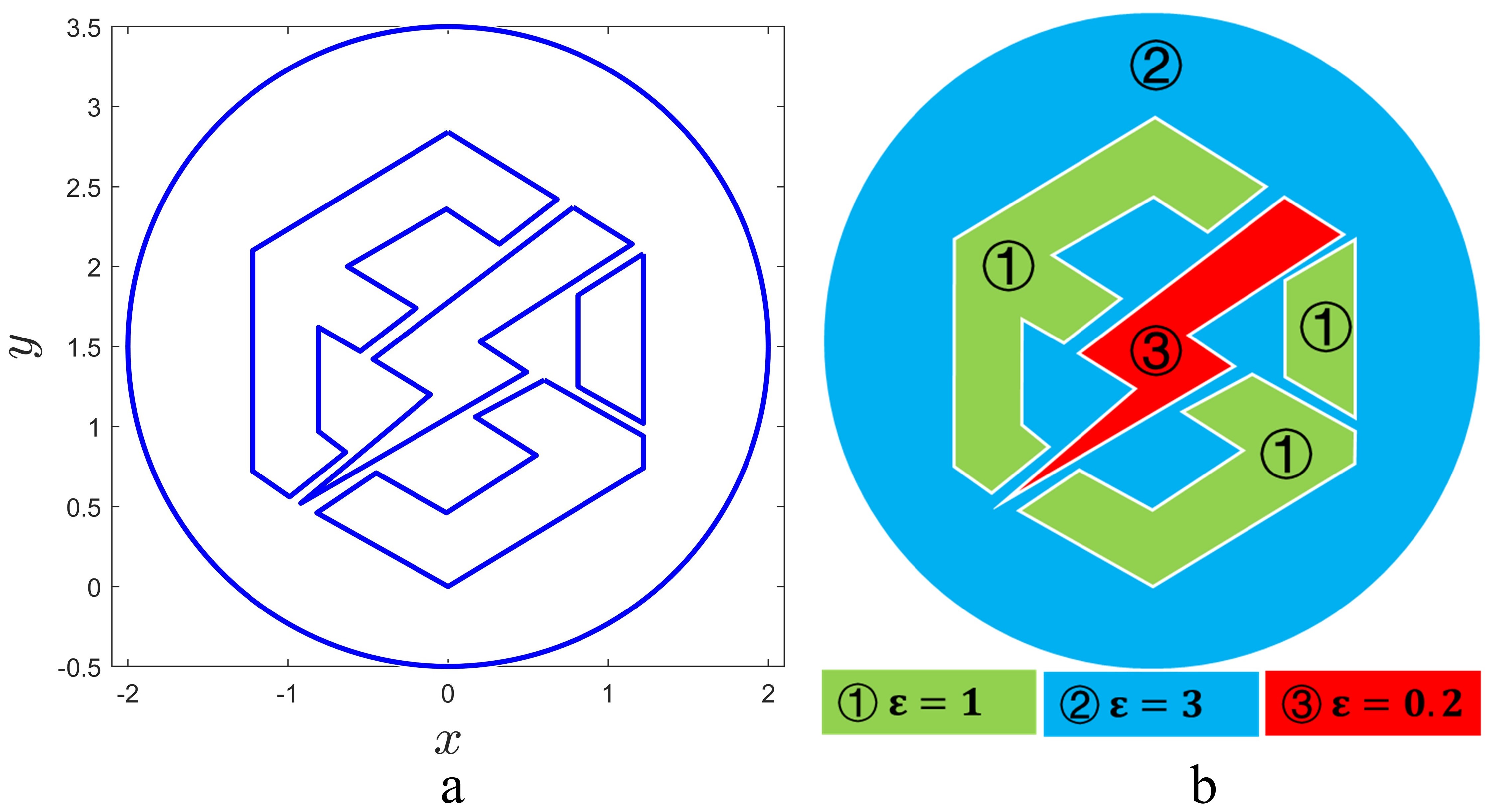}
    \caption{Computation domain and media distribution of example in Section 4. 
    (a) The insignia of the School of Electrical Engineering of Xi'an Jiaotong University is used as the computation domain. 
    (b) The computation domain is divided into three parts filled with different material. Part one marked by color green is fill with material with the permittivity of 1, part two marked by color blue is fill with material with the permittivity of 3 and part three marked by color red is fill with material with the permittivity of 0.2.}
  \end{figure}

  \begin{figure}[!h]
    \centering
    \includegraphics[scale=.75]{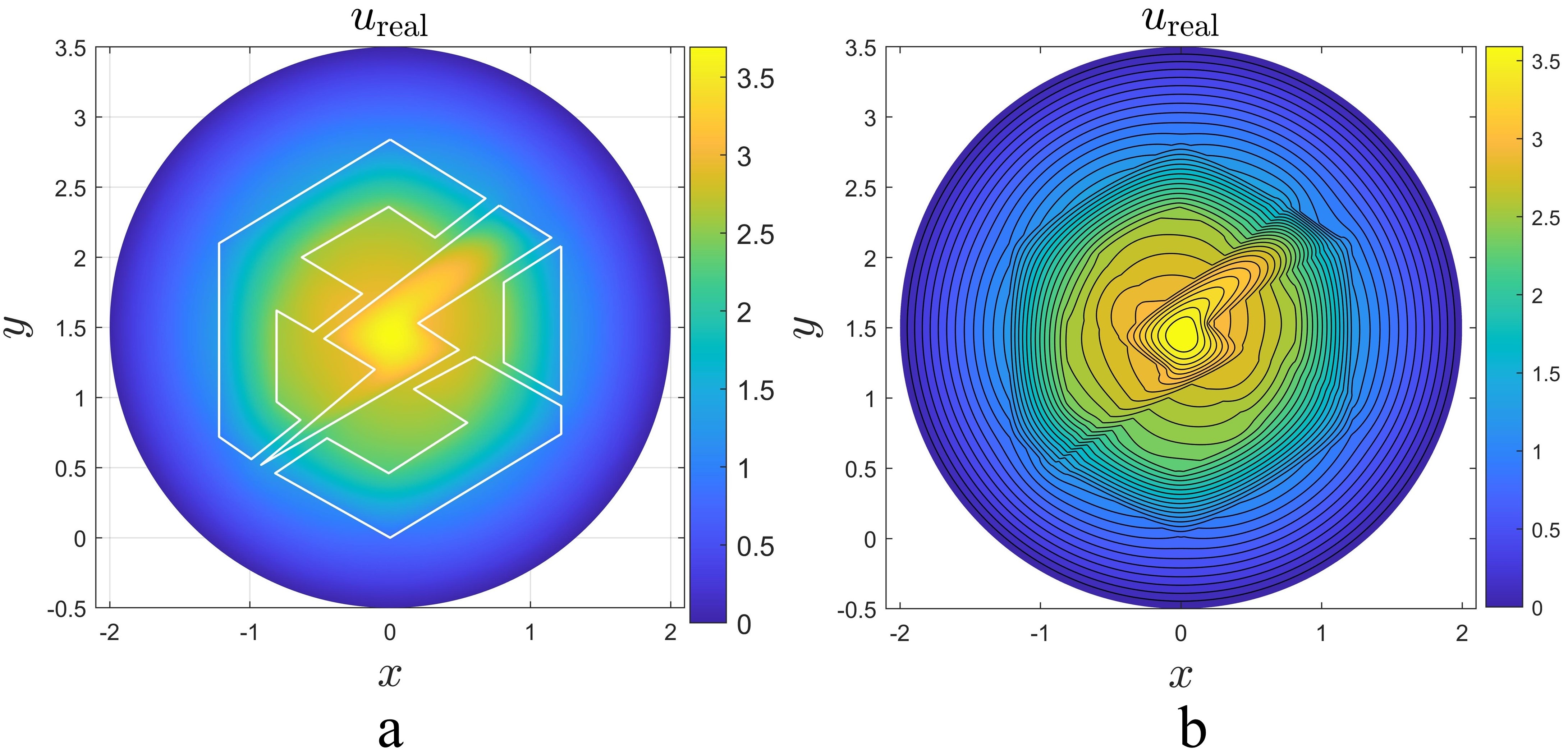}
    \caption{The real solution of example in Section 4. (a) shows the real solution together with interfaces of subdomain marked as white lines. (b) shows the real solution together with its contours.}
  \end{figure}

Among them, $\partial \Omega$ represents the boundary of the solution domain, and its geometric shape is defined by the equation $x^2+(y-1.5)^2=2^2$. 
  For problems involving piecewise homogeneous media and irregular subdomains, it is often impossible to obtain analytical solutions. 
  Therefore, we use the FEM results on a fine mesh as the reference solution. 
  As shown in Fig. 13-a, we use the FEM calculation results under more than 300,000 mesh elements as the benchmark solution. 
  Furthermore, the distribution of the solution's equipotential lines is displayed in Fig. 13-b. 
  First, we used FEM to solve the problem. The results of discretizing the domain utilizing 471 triangular elements are displayed in Fig. 14-a1. 
  Fig. 14-a2 shows the pointwise absolute error distribution between the FEM results and the reference solution, with a maximum value of approximately 0.33 occurring at the center of the solution domain. 
  The standard PINNs method struggles to effectively handle this piecewise homogeneous medium problem.To this end, we adopt conservative physics-informed neural networks (cPINNs) \citep{jagtap2020}.
  \begin{figure}[!t]
    \centering
    \includegraphics[scale=.68]{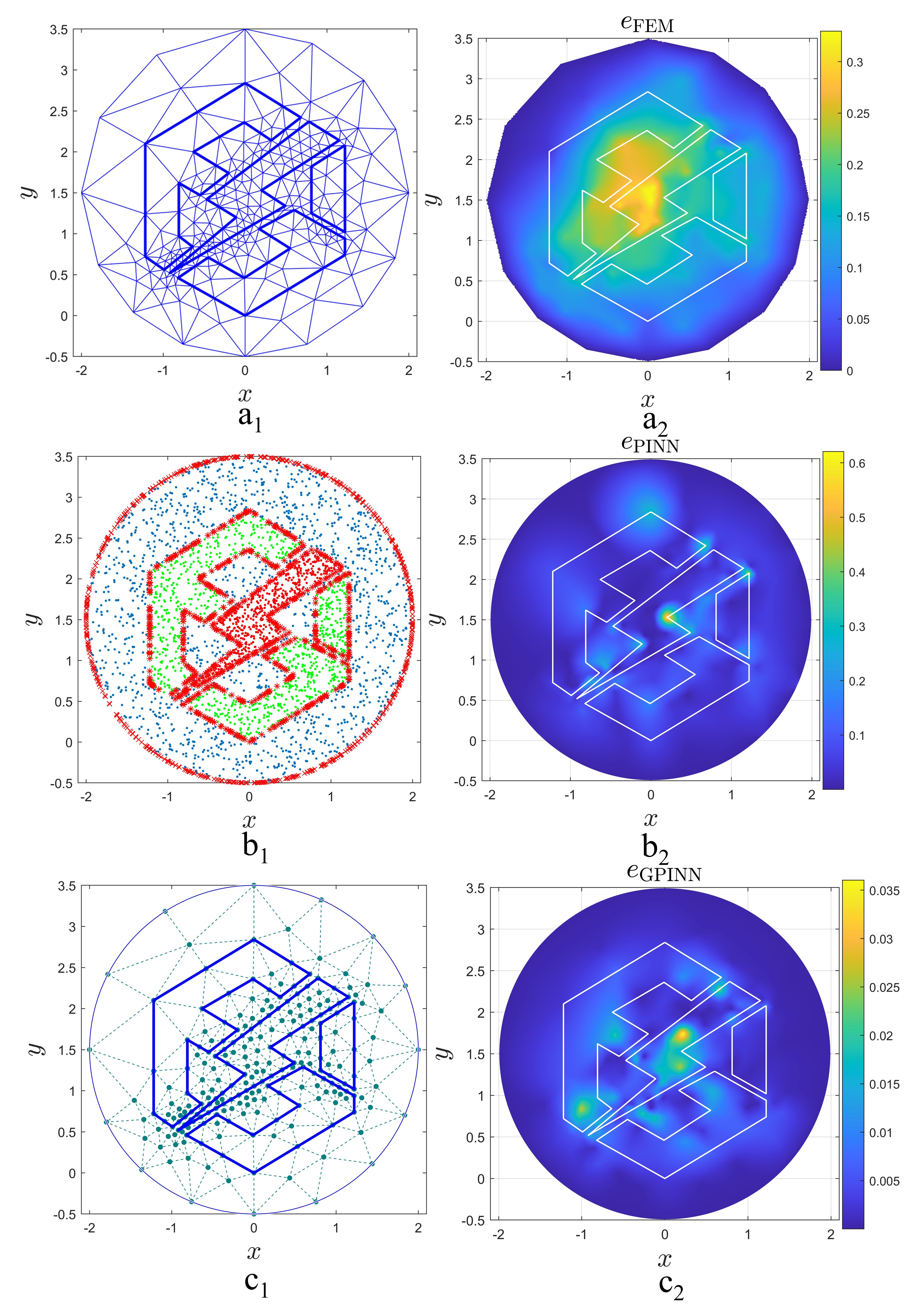}
    \caption{Example in Section 4: Absolute value error under different methodologies. 
    (a1) Domain discretization using the Delaunay triangles.
    (a2) The corresponding absolute value error of FEM. 
    (b1) shows scatter points randomly sampled at each epoch from the boundary marked by the red crosses and interior marked by the green, blue and red points for different subdomains respectively. 
    (b2) The corresponding absolute value error of cPINNs. 
    (c1) Network of nodal association. 
    (c2) The corresponding absolute value error of GPINNs.}
  \end{figure}

\begin{table*}[!h]
  \caption{\label{tab1} Hyperparameters of networks in three subdomains}
  \centering
  \begin{tabular}{p{5.5cm}p{3cm}p{3cm}p{3cm}}
    \hline
    Subdomain number & Subdomain 1 & Subdomain 2 & Subdomain 3 \\
  \hline
  Number of hidden layers        & 3   & 3          &2       \\
  Neurons in each hidden layer   & 50  & 80         &50      \\
  Residual points   &1200 & 1400       &500     \\
  Boundary points   &0    & 2000       &0       \\
  Interface points  &1000 & 1500       &500     \\
  Adaptive activation function   &tanh & tanh       &tanh \\
  \hline
  \end{tabular}
\end{table*}

cPINNs divides the entire computational domain into multiple subdomains, with the solution for each subdomain represented by an independent neural network. 
In this case, we describe the subdomains of three different media using three separate neural networks, tailoring the neural network architecture according to the subdomains' scale and geometric complexity. 
In particular, we select a tiny neural network for the boundary-simple area 3 with the structure of \([2,\,50,\,50,\,1]\). 
We then use a bigger neural network, specifically \([2,\,50,\,50,\,50,\,1]\), for domain 1. 
Lastly, we will use the structure \([2,\,80,\,80,\,80,\,1]\) to further increase the size of the neural network because domain 2 has the largest area and the most complicated boundary. 
Taking subdomain 2 as an example, we increase the number of sampling points on the interface and boundary to 1500 and 2000, respectively, in each epoch to give boundary information and enhance the communication efficiency between neural networks on the corresponding subdomains. 
The total number of epochs is set to 100,000. 
Table 5 provides more detailed information about other hyperparameters of the network. 
Fig. 13-b1 shows the distribution of random sampling nodes, where the red "x" represents boundary sampling points, and the green, blue, and red circles represent sampling points in domains 1, 2, and 3, respectively. 
The activation function in this example is still the hyperbolic tangent function, tanh. 
We apply a scaling factor $A$ to regulate the activation function's slope in accordance with the results presented in the paper \citep{jagtap2020adaptive}. 
There the convergence of the loss function can be accelerated. Initially, $A$ is set to 0.05 and dynamically optimized during the network training process. 
The adaptive hyperbolic tangent activation function can be expressed as follows:

\begin{equation}
  \text{Adaptive tanh}\left(x\right)=\frac{e^{Ax}-e^{-Ax}}{e^{Ax}+e^{-Ax}}
\end{equation}

  Since the physical characteristics of the medium in each subdomain vary and cause corresponding modifications to the governing equations, 
  the loss function of cPINNs is determined by the physical information of each subdomain. 
  Furthermore, the loss function also needs to introduce continuity constraints at the interface, especially the conservation of flux and the continuity of the solution function at the interface. 
  Taking the loss function in domain 2 as an example, the specific expression is as follows:

\begin{equation}
  \mathrm{Loss}(\theta)=w_\mathrm{bou}\times\mathrm{MSE}_\mathrm{bou}+w_\mathrm{pde}\times\mathrm{MSE}_\mathrm{pde}+w_\mathrm{flux}\times\mathrm{MSE}_\mathrm{flux}+w_\mathrm{con}\times\mathrm{MSE}_\mathrm{con}
\end{equation}
where $w_{bou}$ ,$w_{pde}$ ,$w_{flux}$ and $w_{con}$ represent the weights of the residuals for boundary, governing equation, interface flux conservation, and interface solution continuity, respectively. 
We set all these weights to one. 
MSE stands for mean squared error calculated for each term using the following formula:
\begin{equation}
  \begin{aligned}
    &\mathrm{MSE}_{bou}= \frac1{N_B}\sum_{i=1}^{N_B}\Bigl[u_{\theta_2}\Bigl(x_i^B,y_i^B\Bigr)\Bigr]^2 \\
    &\mathrm{MSE}_{pde}= \frac{1}{N_P}\sum_{i=1}^{N_P}\Bigl[-\nabla\cdot 3\times\nabla u_{\theta_2}\Bigl(x_i^P,y_i^P\Bigr)-6\Bigr]^2 \\
    &\mathrm{MSE}_{flux}= \frac{1}{N_I}\sum_{i=1}^{N_I}\Bigl[\nabla u_{\theta_2}\Bigl(x_i^I,y_i^I\Bigr)\cdot \mathbf{n}-\nabla u_{\theta_j}\Bigl(x_i^I,y_i^I\Bigr)\cdot \mathbf{n}\Bigr]^2 \\
    &\mathrm{MSE}_{con}= \frac{1}{N_I}\sum_{i=1}^{N_I}\biggl[u_{\theta_2}\left(x_i^I,y_i^I\right)-\frac{1}{2}\biggl(u_{\theta_2}\left(x_i^I,y_i^I\right)+u_{\theta_j}\left(x_i^I,y_i^I\right)\biggr)\biggr]^2 
    \end{aligned}
\end{equation}

where \( u_{\theta_j}(x,y), (j=1,3) \) respectively represent the neural network functions in adjacent subdomains 1 or 3, 
$\Bigl(x_i^B,y_i^B\Bigr)$, $\Bigl(x_i^P,y_i^P\Bigr)$, $\Bigl(x_i^I,y_i^I\Bigr)$ represent the sampling points on the boundary, the solution domain, and the material interface, respectively. 
$N_B$, $N_P$, and $N_I$ represent the number of boundary points, interior points in domain 2, and sampling points on the interface, respectively. 
Even with extensive and meticulous preparation, the results of cPINNs are still unsatisfactory. 
Due to the rapid gradient change at the sharp corners of the geometric boundary, the maximum error occurs, as shown in Fig. 13(b2). 
By incorporating the edge feature information of the node topology network, GPINNs can more effectively correct the errors on the topology nodes, thereby improving accuracy, as seen in Fig. 13(c1). 
Fig. 13 (c2) displays the distribution of the absolute error of GPINNs' prediction results.

\begin{table*}[!h]
  \caption{\label{tab1} Results of the different solving methods}
  \centering
  \begin{tabular}{p{2cm}p{4.5cm}p{4.3cm}p{3.2cm}}
    \hline
   Type & Number of discrete nodes & Maximum absolute error & $L_2$ norm error \\
  \hline
  FEM       & 244 & 0.3304   &0.1812  \\
  cPINNs   & 244 & 0.6205   &0.0904  \\
  GPINNs & 244 & 0.0368   &0.0070  \\
  \hline
  \end{tabular}
\end{table*}

The detailed results of FEM, cPINNs, and GPINNs are shown in Table 5. 
GPINNs greatly lowers the $\mathrm{L}_2$ norm error and the maximum absolute error, by almost an order of magnitude, when compared to FEM and cPINNs. 
This result further confirms the effectiveness of GPINNs in solving the piecewise homogeneous medium problem. 
However, the accuracy improvement of GPINNs in this case is substantially diminished, declining by approximately an order of magnitude when compared to previous cases. 
The primary reason of this issue is that excessive error oscillations are introduced at the interface by cPINNs, which has a detrimental effect on the error's general homogeneity. 
Actually, it is not an unusual phenomena for PINNs to exhibit error oscillations at low-regularity areas like interfaces or borders. 
Numerous researchers have extensively studied this issue and proposed various improvement methods to enhance the training performance of PINNs. 
These techniques include the Residual-based Adaptive Refinement (RAR) method \citep{lu2021deepxde}, 
importance sampling methods \citep{nabian2021efficient}, and their improved versions, 
including Residual-based Adaptive Distribution (RAD) and Residual-based Adaptive Refinement Distribution (RAR-D) \citep{wu2023comprehensive}. 
However, research on employing PINNs and its variations to solve the piecewise homogeneous media issues has not advanced very quickly thus far. 
Consequently, we think that a more thorough study of the fundamental principles behind the piecewise homogeneous media problem and to build algorithmic architectures that can adhere to the laws of physical conservation are necessary. 
The GPINNs method, which is based on conservation-type node topology networks, may offer a viable solution to the piecewise homogeneous media problem.

\section{Conclusion}

In summary, this paper proposes a Global physics-informed neural networks (GPINNs)  based on a conservation-formulated node network. 
The global correlation strategy of randomly distributed nodes not only ensures the uniqueness of the solution but also exhibits greater flexibility and higher accuracy in solving problems with complex boundaries and piecewise homogeneous media. 
Compared to PINNs, which enforce pointwise PDE constraints, GPINNs demonstrate a significantly faster convergence rate. 
Furthermore, in both one-dimensional and two-dimensional cases, GPINNs achieve accuracy improvements of \(3-4\) and \(2-3\) orders of magnitude, respectively, compared to traditional spatially discretized numerical methods. 
Additionally, the proposed method leverages a mesh-free representation with randomly distributed nodes, eliminating the grid dependency inherent in conventional numerical approaches. 
This characteristic makes GPINNs particularly suitable for solving high-dimensional and complex boundary problems. 

Moreover, It is important to note that the accuracy of the prior knowledge used to estimate node correlations plays a critical role in determining the overall solution accuracy of GPINNs. 
A key challenge for further improving GPINNs accuracy lies in obtaining effective prior knowledge, especially for problems with low-regularity solutions. 
Addressing this challenge will be a primary focus of future research.

\section*{Acknowledgments}

\bibliographystyle{model1-num-names}
\bibliography{refs618.bib}

\end{document}